\documentclass[11pt]{article}
\usepackage{arxiv}

\usepackage[utf8]{inputenc}
\usepackage[T1]{fontenc}
\usepackage{lmodern}
\usepackage[protrusion=true,expansion=true]{microtype} 
\usepackage{graphicx}
\usepackage{booktabs}
\usepackage{multirow}
\usepackage{amsmath,amssymb,amsfonts,amsthm,mathtools,bm}
\usepackage{enumitem}
\usepackage{array}
\usepackage{url}
\usepackage{hyperref}
\usepackage{cleveref}
\usepackage{xcolor}
\usepackage{doi}
\usepackage{natbib}
\usepackage{setspace}
\usepackage{appendix}
\usepackage{geometry}
\hypersetup{colorlinks=true,linkcolor=blue,citecolor=blue,urlcolor=blue}
\usepackage{tikz}
\usetikzlibrary{positioning,fit,arrows.meta,calc}
\newtheorem{theorem}{Theorem}
\newtheorem{proposition}[theorem]{Proposition}
\newtheorem{lemma}[theorem]{Lemma}
\newtheorem{corollary}[theorem]{Corollary}
\theoremstyle{definition}
\newtheorem{definition}[theorem]{Definition}

\theoremstyle{remark}
\newtheorem{remark}[theorem]{Remark}

\newcommand{\R}{\mathbb{R}}

\newcommand{\norm}[1]{\left\lVert #1 \right\rVert}
\newcommand{\inner}[2]{\left\langle #1,#2 \right\rangle}
\newcommand{\Had}{\odot}
\newcommand{\diag}{\operatorname{diag}}
\newcommand{\blkdiag}{\operatorname{blkdiag}}

\title{A Co-Evolutionary Theory of Human–AI Coexistence: Mutualism, Governance, and Dynamics in Complex Societies}

\author{
  Somyajit Chakraborty \\
  School of Chemistry and Chemical Engineering, \\
  Shanghai Jiao Tong University, Shanghai 200240, China \\
  \texttt{chksomyajit@sjtu.edu.cn}
}
\date{}

\begin{document}
\maketitle

\begin{abstract}
Classical robot ethics is often framed around obedience, including Asimov's laws. This framing is insufficient for contemporary AI systems, which are increasingly adaptive, generative, embodied, and embedded in physical, psychological, and social environments. This paper proposes \emph{conditional mutualism under governance} as a framework for human--AI coexistence: a co-evolutionary relationship in which humans and AI systems develop, specialize, and coordinate under institutional conditions that preserve reciprocity, reversibility, psychological safety, and social legitimacy. We synthesize work from computability, automata theory, statistical machine learning, neural networks, deep learning, transformers, foundation models, world models, embodied AI, alignment, human--robot interaction, ecological mutualism, biological markets, coevolution, and polycentric governance. We formalize coexistence as a multiplex dynamical system across physical, psychological, and social layers, incorporating reciprocal supply--demand coupling, conflict penalties, developmental freedom, and governance regularization. The framework yields conditions for existence, uniqueness, and global asymptotic stability of coexistence equilibria. We complement the analytical results with deterministic ODE simulations, basin-of-attraction sweeps, sensitivity analyses, governance-regime comparisons, shock tests, and local stability checks. The simulations indicate that governed mutualism reaches a high coexistence index with negligible domination, whereas insufficient or excessive governance can produce domination, weak-benefit lock-in, or suppressed developmental freedom. The results suggest that reciprocal complementarity can support stable coexistence, while ungoverned coupling can increase fragility, lock-in, polarization, and domination risk. Human--AI coexistence should therefore be treated as a co-evolutionary governance problem rather than a static obedience problem. This reframing provides an analytical basis for coexistence principles that permit bounded AI development while preserving human dignity, contestability, collective safety, and fair distribution of gains.
\end{abstract}

\section{Introduction}
Artificial intelligence has evolved through several distinct but cumulative intellectual regimes, and this historical development is relevant to any theory of human--AI coexistence. The earliest regime was formal and symbolic. Computability theory, finite automata, cybernetics, and early symbolic AI established the machine as a rule-governed artifact that could process symbols, execute procedures, and realize bounded forms of reasoning \citep{turing1937,rabin1959,mcculloch1943logical,shannon1948mathematical,wiener1948cybernetics,newell1956logic,mccarthy1959programs,nilsson1980principles}. Within this paradigm, the machine was commonly conceptualized as an instrument: it executed specified instructions, and its relation to human purposes could be represented as a hierarchy of command, specification, and compliance. This framing continues to influence public and policy discourse on artificial intelligence, although it is increasingly insufficient for systems currently deployed or under development.

A second regime emerged with statistical learning and connectionism. Rather than treating intelligence as the explicit execution of hand-written symbolic rules, statistical learning theory reframed the problem around generalization from finite data under uncertainty \citep{vapnik1999nature,mitchell1997machine,bishop2006pattern,jordan2015mltrends}. In parallel, neural and connectionist models showed that useful internal representations could emerge from distributed adaptive systems rather than from transparent symbolic programming alone \citep{rosenblatt1958perceptron,hopfield1982neural,rumelhart1986learning,bengio1994longterm,hochreiter1997lstm}. This shift is conceptually important for the present paper because it reframes AI as less a static executor of rules and more a learner whose behavior depends on data, architecture, inductive bias, and interaction history. When systems learn from data rather than only execute explicit rules, the question of coexistence can no longer be reduced to unidirectional command relations.

A third regime, consolidated by deep learning, sequence modeling, attention, and large-scale generative training, shifted AI from task-specific estimators toward general-purpose representational infrastructures \citep{lecun2015deep,goodfellow2016deep,schmidhuber2015deep,krizhevsky2012imagenet,sutskever2014seq2seq,cho2014rnn,bahdanau2015neural,vaswani2017attention}. Transformer-based systems, self-supervised objectives, and scaling laws enabled models that are predictive, compositional, reusable, and increasingly multimodal \citep{devlin2019bert,raffel2020t5,brown2020language,kaplan2020scaling,hoffmann2022chinchilla,chowdhery2022palm,touvron2023llama}. Generative modeling extended this development by enabling models to synthesize text, images, video, and action-relevant trajectories rather than merely classify inputs \citep{kingma2013auto,rezende2014stochastic,goodfellow2014gan,sohl2015deepunsup,ho2020ddpm,song2021score,rombach2022ldm}. The foundation-model paradigm then crystallized the idea that a single pre-trained model can become a widely reused substrate for downstream systems, social practices, and institutional workflows \citep{bommasani2021foundation,weidinger2022taxonomy,bubeck2023sparks,openai2023gpt4}. In this setting, the relevant unit of analysis is no longer a narrow tool, but an adaptive and highly reusable model ecosystem.

\subsection{From Foundation Models to Physical AI}
The present landscape adds a fourth regime: world modeling, simulation-based learning, and physical or embodied AI. Classical model-based reinforcement learning already suggested that intelligence benefits from internal predictive models of the environment \citep{sutton1991dyna,kaelbling1996reinforcement,deisenroth2011pilco,nagabandi2018neural}. Contemporary world-model work makes that intuition more general and technically central. The original \emph{World Models} program proposed compact latent models in which policies could be trained through imagination \citep{ha2018world}. The Dreamer line and related latent-dynamics methods then showed that planning and control can be carried out directly in learned latent state spaces with improved efficiency and generality \citep{hafner2019dream,hafner2020dream,hafner2023mastering,sekar2020planning,hansen2022tdmpc,micheli2023transformerwm}. More recent work extends the concept from reinforcement learning into broader questions of representation, simulation, and internal structure, including evidence that transformer systems can encode nontrivial spatiotemporal and causal structure that exhibits increasingly world-model-like structure \citep{gurnee2024space,spies2024causalworldmodels}. This matters for coexistence because the central issue is no longer limited to output correctness. It also concerns systems that can learn internal models of environments, represent possible futures, and select actions on the basis of those representations.

At the same time, the boundary between foundation models and embodied systems is becoming less distinct. Vision-language-action systems now connect language grounding, perception, and action in ways that bring large models into direct contact with the physical world \citep{driess2023palme,brohan2022rt1,brohan2023rt2,reed2022gato,ahn2022saycan,openx2024,octo2024,openvla2024,chi2023diffusionpolicy}. Recent work on physical AI and world foundation models extends this direction by treating video-rich, multimodal simulation as a substrate for training agents that will later operate in real environments \citep{liu2025cosmos,fung2025embodied}. The field is therefore moving from symbolic and disembodied reasoning toward model-based, multimodal, and eventually physically situated intelligence. When action is included in the system loop, runtime governance, intervention, and reversibility become engineering requirements rather than optional policy additions.

\begin{figure}[t]
    \centering
    \includegraphics[width=\linewidth]{Figures/Figure1.png}
    \caption{Evolution of AI paradigms and the corresponding shift in the governance problem, from rule specification to runtime-constrained coexistence.}
    \label{fig:ai_evolution_governance}
\end{figure}

Figure~\ref{fig:ai_evolution_governance} summarizes how the ontology of AI expands across paradigms and why governance must move from static rule-setting toward runtime-aware coexistence. This landscape further clarifies the alignment problem. Obedience to surface instructions is no longer enough when human preferences are incomplete, under-specified, conflicting, strategically expressed, or only partially observable. Modern alignment research already reflects this point. Cooperative inverse reinforcement learning treats human--AI interaction as a cooperative process under uncertainty about the true objective \citep{hadfield2016cirl}. Corrigibility and interruptibility research further argues that useful systems must remain deferential, updateable, and governable rather than blindly persistent in pursuit of fixed goals \citep{soares2015corrigibility,amodei2016concrete,russell2015researchpriorities}. Preference learning, instruction tuning, RLHF, and constitutional or rule-guided alignment improve useful behavior but do not eliminate the broader problem that AI systems now participate in social worlds where multiple stakeholders, institutions, and normative constraints interact \citep{christiano2017preference,ouyang2022instructgpt,bai2022constitutional,jobin2019global,cath2018governance,floridi2018ai4people}. These developments motivate renewed analysis of human--AI coexistence: the field has moved beyond the conceptual simplicity of a single obedient machine serving a single user.

\subsection{Motivation for a Coexistence Framework}
Two gaps motivate this paper. First, the dominant public template for thinking about AI coexistence remains obedience-centered. Asimov's laws remain culturally influential, but they presume a static hierarchy in which the machine is fundamentally a servant and harm is reducible to a relatively simple imperative \citep{asimov1950irobot}. That framing is poorly suited to a world of adaptive foundation models, world-model-based agents, embodied systems, and institutional ecosystems in which both humans and AI systems shape each other's development. Second, existing technical and governance literatures are often siloed. Alignment theory addresses objective uncertainty and assistance \citep{hadfield2016cirl,russell2019humancompatible}; HRI studies examine trust and social perception \citep{hancock2011meta,degraaf2019}; governance frameworks emphasize accountability and risk management \citep{oecd2019ai,nist2023airmf,euaiact2024}; ecological theory offers formal intuitions about stable mutualism and bounded exploitation \citep{noe1994biological,bronstein2015mutualism,holland2002consumer}. What remains underdeveloped is a unified framework that treats coexistence as a \emph{co-evolutionary, multi-layer, and mathematically tractable} problem.

\subsection{Objectives and research questions}

This paper develops such a framework. It is organized around four research questions:

\begin{enumerate}[label=\textbf{RQ\arabic*.},leftmargin=2em]
    \item How should human--AI coexistence be conceptualized once AI systems become generative, foundation-based, world-model-driven, and increasingly embodied?

    \item Why are obedience-centered laws insufficient for physical, psychological, and social coexistence, and how does conditional mutualism offer a better conceptual frame?

    \item What can ecology, biological markets, coevolution, and mutualism teach us about durable coexistence, and how can these insights be translated into a formal dynamical model?

    \item Under what conditions does the resulting model admit stable coexistence equilibria, and what design principles follow for future governance, design, and evaluation?
\end{enumerate}

\subsection{Methodological Strategy and Contributions}
Methodologically, the paper combines interdisciplinary synthesis with formal theory-building. We first integrate technical AI history, recent world-model and embodied-agent literature, psychological and sociotechnical findings, and ecological theories of coexistence. We then formalize human--AI coexistence as a multiplex dynamical system on physical, psychological, and social layers with reciprocal supply--demand coupling and governance as a stabilizing control term. The paper makes five main contributions. First, it reframes coexistence from obedience to conditional mutualism under governance. Second, it develops an integrated review connecting foundational AI, world models, HRI, alignment, ecology, and governance. Third, it proposes a mathematical framework with lemmas, propositions, and theorems establishing boundedness, existence, uniqueness, and stability under explicit assumptions. Fourth, it tests the model through deterministic simulations, basin maps, sensitivity analyses, governance-regime comparisons, shock-resilience experiments, and equilibrium stability checks. Fifth, it uses the formal and numerical results to articulate design principles for coexistence centered on reciprocity, bounded autonomy, reversibility, psychological integrity, and institutional contestability.

\subsection{Roadmap}
The rest of the paper proceeds as follows. Section~\ref{sec:related} reviews the relevant literature in detail. Section~\ref{sec:conceptual} develops the conceptual shift from Asimovian obedience to conditional mutualism under governance. Section~\ref{sec:math} introduces the multiplex dynamical system and coexistence functional. Section~\ref{sec:results} states and proves the main mathematical results. Section~\ref{sec:numerical} reports numerical simulations, basin maps, sensitivity analyses, shock tests, and stability checks. Section~\ref{sec:discussion} discusses implications for world models, embodied AI, general-purpose AI systems, governance, and future empirical work. Section~\ref{sec:conclusion} concludes.

\section{Related Work}\label{sec:related}

\subsection{From formal computation to foundation models}
A rigorous theory of human--AI coexistence should account for how the ontology of AI has changed across the history of the field. The earliest computational paradigm treated machine intelligence as formal procedure. Turing's account of effective computability, the theory of automata, cybernetics, and early symbolic AI all portrayed the machine as an entity whose competence arose from explicit procedure, symbolic representation, and bounded state transition \citep{turing1937,rabin1959,mcculloch1943logical,shannon1948mathematical,wiener1948cybernetics,newell1956logic,mccarthy1959programs}. This paradigm remains relevant because it explains why obedience-based thinking is so intuitive: if a machine is fundamentally a rule-executing artifact, then control appears to be a matter of writing the right rule set. However, modern AI no longer fits this description except in residual or hybrid form.

Statistical learning introduced a different picture. In this setting, the central problem became not hand-crafted reasoning but inference and generalization from finite samples under uncertainty \citep{vapnik1999nature,mitchell1997machine,bishop2006pattern,jordan2015mltrends}. In this frame, the machine is not only specified through instructions. It is trained to infer patterns, compress structure, and make predictions from data. This move already complicates the normative and institutional analysis of AI because system behavior depends not only on programmer intent but also on data distributions, objective functions, model class, and deployment context. Intelligence is better characterized as adaptive statistical behavior than as obedience alone.

Connectionist and neural-network traditions deepened that shift by showing that rich internal structure could arise through learning rather than explicit symbolic programming \citep{rosenblatt1958perceptron,hopfield1982neural,rumelhart1986learning,bengio1994longterm,hochreiter1997lstm}. Deep learning then made hierarchical representation learning the dominant engineering paradigm across perception, language, and control \citep{lecun2015deep,goodfellow2016deep,schmidhuber2015deep,krizhevsky2012imagenet}. Sequence-to-sequence learning and neural attention moved the field closer to flexible conditional generation and cross-modal mapping \citep{sutskever2014seq2seq,cho2014rnn,bahdanau2015neural}. This phase is important for the present paper because it marks the point at which AI systems are better characterized as adaptable representational systems than as fixed software modules, with internal states that are partially learned, partially opaque, and often emergent.

Transformers substantially accelerated that transition. Attention-based architectures made large-scale sequence modeling more parallel, transferable, and general \citep{vaswani2017attention}. Subsequent systems such as BERT, T5, GPT-style language models, and multimodal pretraining architectures demonstrated that a single model family can support a wide range of tasks via self-supervised pretraining and lightweight adaptation \citep{devlin2019bert,raffel2020t5,brown2020language,radford2021clip,alayrac2022flamingo}. Scaling-law research and large-scale empirical studies suggested that capability is not merely a function of hand-designed modularity, but of data scale, parameter scale, and compute allocation \citep{kaplan2020scaling,hoffmann2022chinchilla,chowdhery2022palm,wei2022emergent}. The implication for coexistence is important: the socially relevant unit is no longer a narrowly scoped expert system but a broadly capable substrate that can be reused across domains, users, and institutions.

Generative modeling further shifted AI from task-specific prediction toward synthetic content and trajectory generation. Variational autoencoders, adversarial networks, diffusion models, and latent diffusion systems made it possible for AI to synthesize novel candidate texts, images, videos, and trajectories \citep{kingma2013auto,rezende2014stochastic,goodfellow2014gan,sohl2015deepunsup,ho2020ddpm,song2021score,rombach2022ldm}. This matters not only because generation is practically important, but because coexistence with generative systems is inherently prospective: these systems help shape the informational and practical futures humans then inhabit. They do not only respond to existing information environments; they increasingly influence the informational and practical contexts in which subsequent decisions occur.

The foundation-model paradigm made the consequences even more systemic. A single pre-trained model can now serve as infrastructure for numerous applications, organizations, and downstream agentic pipelines \citep{bommasani2021foundation,weidinger2022taxonomy,openai2023gpt4,bubeck2023sparks,touvron2023llama}. This infrastructural role changes the coexistence question from ``how should one user control one machine?'' to ``how should societies govern highly reusable adaptive systems whose capabilities propagate across many contexts?'' The remainder of this paper takes this transformed ontology as its starting point.

\subsection{World models, simulation, and the advent of physical AI}
If the transformer/foundation-model turn expanded the scale and reuse of AI, the world-model turn changes its temporal and causal orientation. A world model, in the broad sense relevant here, is an internal model that enables an agent to represent aspects of its environment, anticipate the consequences of actions, and support policy selection under uncertainty. This idea has older roots in model-based reinforcement learning and adaptive control \citep{sutton1991dyna,kaelbling1996reinforcement,deisenroth2011pilco,nagabandi2018neural}, but recent work has made the concept increasingly central to mainstream AI. The original \emph{World Models} paper showed that compact latent models of dynamics can support learned control through imagined rollouts \citep{ha2018world}. The Dreamer family developed this further by learning latent dynamics models that support long-horizon control and planning in imagination without requiring explicit symbolic simulation \citep{hafner2019dream,hafner2020dream,hafner2023mastering}. Related methods such as latent planning and temporal-difference world-model control further demonstrate that predictive latent structure can be used not merely for passive modeling, but for active decision making \citep{sekar2020planning,hansen2022tdmpc,micheli2023transformerwm}.

World models are important for coexistence because they move AI from reactive pattern matching toward anticipatory agency. A system with an internal model of consequences is a system that can generate futures, compare them, and act conditionally on imagined outcomes. This changes both safety and governance. The challenge is no longer only whether a model's immediate output is acceptable, but whether the internal simulation and planning process that precedes action is calibrated, governable, and aligned with legitimate human and institutional constraints. Recent work on internal structure in large models reinforces this point. Evidence that language models can encode coherent spatial and temporal structure, and that transformers in structured domains can build causally meaningful internal representations, suggests that the boundary between language modeling and world modeling is becoming increasingly porous \citep{gurnee2024space,spies2024causalworldmodels}. Even when these models are not full-fledged physical simulators, they may contain partial world models sufficient to affect planning, coordination, and decision support.

A related development is embodiment. Once predictive internal models are linked to perception and action, the world-model question becomes inseparable from robotics and physical AI. PaLM-E, RT-1, RT-2, Gato, SayCan, Diffusion Policy, Open X-Embodiment, Octo, and OpenVLA all indicate a broad shift toward generalist, language-conditioned, or internet-scale embodied systems that couple vision, language, action, and policy learning \citep{reed2022gato,ahn2022saycan,driess2023palme,brohan2022rt1,brohan2023rt2,chi2023diffusionpolicy,openx2024,octo2024,openvla2024}. These systems differ in capability and maturity, but collectively they signal a transition from disembodied reasoning engines to agents that can interpret instructions, map them onto sensory input, and produce real or quasi-real actions in the world.

At this point, the older obedience paradigm becomes limited. In embodied systems, action unfolds in real time under partial observability, uncertainty, shifting environmental conditions, and often delayed or sparse feedback. It is difficult to reduce such systems to a static set of prohibitions or to assume that upstream instruction alignment alone guarantees safe behavior. Runtime governance becomes a central technical requirement because the model's learned policy, world model, and action interface can interact in ways that are difficult to fully specify ex ante. Recent work on embodied AI governance and physical AI world foundations explicitly reflects this shift by arguing that digital twins, simulation infrastructures, action constraints, and post-training or deployment guardrails are indispensable components of future AI stacks \citep{perlo2025embodied,liu2025cosmos,fung2025embodied}. Thus, when agents act through world models in physical environments, governance must move inside the loop.

The emergence of world foundation models strengthens this conclusion. Recent systems treat world modeling not as a narrow robotics subproblem but as a general-purpose modeling substrate that can be post-trained or adapted for downstream physical tasks \citep{liu2025cosmos}. This is conceptually analogous to what foundation models did for language and multimodal processing. A world foundation model promises broad reuse, transfer, and simulation leverage, but it also inherits the systemic concerns of foundation models while adding the material risks of physical interaction. A coexistence theory that ignores this development would remain limited to a pre-embodiment picture of AI.

For the present paper, the main implication is that world models change how AI should be analyzed. They introduce anticipation, counterfactual reasoning, temporal depth, and action orientation into the coexistence problem. Human--AI coexistence must therefore be formulated not only as a question of static outputs or isolated decisions, but as a question of coupled dynamical systems in which internal models, human responses, institutions, and material environments evolve together.

\subsection{From foundation models to embodied foundation agents}
A further development is the convergence between foundation modeling and agentic embodiment. Foundation models were initially discussed primarily in relation to language and internet-scale multimodal understanding, but their broader significance lies in their generality, transferability, and infrastructural role \citep{bommasani2021foundation,weidinger2022taxonomy}. Once those properties are connected to action, the result is no longer just a model that answers questions or generates media. It is an embodied foundation agent: a broadly pretrained system that can interpret heterogeneous inputs, maintain temporally extended context, and produce action-relevant outputs in a physical or interactive environment. In that sense, PaLM-E, RT-2, Open X-Embodiment, Octo, and OpenVLA should not be interpreted only as isolated robotics systems. They are early instances of a more general shift toward foundation agents \citep{driess2023palme,brohan2023rt2,openx2024,octo2024,openvla2024}.

This convergence matters for the paper's argument because it changes the scale at which coexistence must be analyzed. The unit of governance is no longer only a single robot or a single model checkpoint. It is a sociotechnical stack: pretraining data, world knowledge, action interfaces, post-training alignment, deployment policies, runtime monitors, human operators, and institutional rules. The more foundation-like an embodied agent becomes, the more downstream contexts it can enter and the more path dependence it can create. This motivates a formulation of coexistence across physical, psychological, and social worlds simultaneously.

\subsection{Alignment, corrigibility, and the limits of obedience}
The alignment literature already contains elements of a post-obedience perspective, even if it does not always use that language. Cooperative inverse reinforcement learning is important because it reframes the human--AI relationship as a cooperative game under uncertainty about the human objective \citep{hadfield2016cirl}. This is an important departure from naive obedience. The system should not merely execute the most literal command it receives; it should instead reason under uncertainty about the underlying objective of the human or relevant set of humans. Russell's control-oriented argument develops the same point at a broader level by emphasizing beneficial assistance, uncertainty about objectives, and the design of systems that remain amenable to correction and shutdown \citep{russell2019humancompatible,russell2015researchpriorities}. Corrigibility work makes the technical stakes explicit: a useful advanced system must remain interruptible, modifiable, and non-adversarial toward external correction \citep{soares2015corrigibility,amodei2016concrete}.

Subsequent alignment practice broadened the engineering toolkit without resolving the broader normative and institutional issue. Preference learning, RLHF, instruction tuning, constitutional AI, and related alignment-by-feedback approaches have improved practical controllability and usability \citep{christiano2017preference,ouyang2022instructgpt,bai2022constitutional}. Yet these methods mostly operate within a framework in which a model is optimized to better reflect a desired response distribution. They do not by themselves solve the problem of plural and contested preferences, institutional asymmetry, or cross-domain spillovers once the same model is deployed across heterogeneous settings. Thus, modern alignment methods improve behavior, but they do not eliminate the need for a broader coexistence framework.

This limitation becomes even clearer for embodied and world-model-based systems. A policy can be instruction-following at the interface level while still producing problematic real-world trajectories under distribution shift, delayed consequences, or incomplete observability. For that reason, recent embodied-agent work increasingly emphasizes runtime monitoring, staged deployment, rollback, and operational guardrails rather than relying on upstream alignment alone \citep{perlo2025embodied,fung2025embodied,liu2025cosmos}. This literature aligns naturally with the central thesis of the present paper: once agents can perceive, imagine, and act in the physical world, governability is not merely a legal or organizational concern but a technical design variable.

The alignment literature also intersects directly with the moral psychology of coexistence. If users systematically over-trust a system, anthropomorphize it, or outsource judgement to it inappropriately, then even a partially aligned model can degrade autonomy and decision quality. This is why obedience is an insufficient normative target. A coexistence regime must preserve usefulness while also preserving corrigibility, contestability, reversibility, and psychological integrity. The mathematical framework developed later in the paper treats these not as secondary properties but as state variables and constraints in a coupled social dynamical system.

\subsection{Trust, Anthropomorphism, and Psychological Integrity}
A second cluster of literature explains why coexistence cannot be reduced to physical safety or task performance. Trust in human--robot interaction is shaped by performance, design, context, and user expectations, but the aim is calibrated trust rather than maximal trust \citep{hancock2011meta}. Automation bias research shows how humans commit omission and commission errors when they over-rely on imperfect automation \citep{parasuraman2010complacency}. The literature on social responses to computers and anthropomorphism shows that humans readily attribute personality, mind, and moral salience to machines, often beyond what the systems warrant \citep{nass2000machines,epley2007anthropomorphism,degraaf2019}. Longitudinal companion-style AI deployments may therefore create not only utility, but dependence, manipulation, and identity effects. 

This body of work motivates one of the paper's central claims: \emph{psychological integrity must be treated as a first-order coexistence condition}. A coexistence regime that prevents bodily injury but systematically induces dependence, deskilling, or social misperception is not a stable or desirable equilibrium.

\subsection{Ecology, biological markets, and coevolution}
The natural-science literature provides a strong theoretical basis for analyzing why coexistence emerges and persists. In ecology and evolution, durable coexistence is rarely unconditional. It depends on repeated interaction, reciprocal gain, ecological constraints, partner choice, and sanctions against exploitation. \citet{axelrod1984evolution} established the enduring relevance of reciprocity in repeated strategic interaction. Biological market theory emphasized supply, demand, and partner choice as mechanisms structuring cooperation \citep{noe1994biological}. Work on conditional mutualism and exploitation in mutualistic systems showed that apparently cooperative relationships can shift when incentives or ecological conditions change \citep{bronstein1994conditional,bronstein2015mutualism}. Consumer-resource models make this intuition dynamical by showing how mutualistic benefits and exploitative pressures can jointly shape stable coexistence or collapse \citep{holland2002consumer}. Coevolutionary theory adds another critical insight: interaction outcomes vary across local ecological and institutional environments, implying that stable coexistence is often context-dependent rather than universal \citep{thompson2005mosaic}. Finally, polycentric governance theory explains why adaptive multi-level institutions often outperform centralized single-stage control in complex systems \citep{ostrom2010polycentric}. 

The analogy to human--AI coexistence is not that AI is literally another species. The relevant analogy is structural: stable coexistence is more likely when both parties gain, neither party can externalize all costs indefinitely, and the surrounding environment supports adaptation, sanction, and exit.

\subsection{Governance Architectures and Institutional Scaffolding}
Recent governance frameworks support many of the normative directions argued for here. The OECD AI Principles emphasize human rights, democratic values, and trustworthy deployment \citep{oecd2019ai}. UNESCO's Recommendation on the Ethics of AI extends the frame to dignity, wellbeing, diversity, and environmental concerns \citep{unesco2021recommendation}. The NIST AI Risk Management Framework treats AI governance as lifecycle risk management \citep{nist2023airmf}. The EU AI Act operationalizes a tiered legal structure in which requirements intensify with risk and domain \citep{euaiact2024}. These documents do not by themselves provide a theory of coexistence, but they collectively show that contemporary governance is already moving beyond simple command-and-control narratives toward risk-sensitive, institutionally layered, and socially embedded forms of oversight.

\subsection{Gap summary}
The literature therefore provides many ingredients of a coexistence theory, but not yet an integrated synthesis. AI history explains why the ontology of the agent has changed. World models and embodied AI explain why physical action and imagined futures matter. HRI and behavioral science explain why trust and manipulation matter. Ecology explains why coexistence requires reciprocity, constraints, and adaptation. Governance scholarship explains why institutions matter. What remains underdeveloped is a single framework that integrates these strands conceptually and mathematically.

\section{From Asimovian Obedience to Conditional Mutualism Under Governance}\label{sec:conceptual}
Asimov's laws remain culturally influential because they offer a simplified normative model: the robot exists as a servant, it must not harm humans, it must obey human orders, and it must preserve itself only within the limits of those higher duties \citep{asimov1950irobot}. This picture is useful as speculative fiction, but it is insufficient for contemporary AI. It assumes a relatively stable hierarchy between a human commander and a machine executor. It also assumes that harm, obedience, and self-preservation can be separated into separable rules. Modern AI systems are not well described by this model. They are trained on large-scale data, reused across many downstream contexts, embedded in institutions, and increasingly connected to planning, simulation, perception, and action. Their effects are therefore not limited to single commands or isolated outputs. They shape information environments, institutional workflows, user expectations, labor structures, and eventually physical action.

The limitation of obedience becomes clearer once one moves from symbolic machines to foundation models and embodied agents. A rule-following artifact can be governed primarily through specification. A learned system must also be governed through data, objectives, feedback, deployment conditions, monitoring, and correction. A physically situated AI system adds another layer: it acts under uncertainty, receives delayed feedback, and may affect the material world before a human supervisor can fully evaluate the consequence. Alignment research already captures part of this shift. Cooperative inverse reinforcement learning treats human--AI interaction as a cooperative process under uncertainty about the true human objective rather than as literal command execution \citep{hadfield2016cirl}. Work on AI safety and corrigibility similarly argues that useful systems must remain interruptible, modifiable, and responsive to correction rather than blindly persistent in pursuit of fixed goals \citep{russell2015researchpriorities,soares2015corrigibility,amodei2016concrete,russell2019humancompatible}. Thus, the problem is not simply that future AI may disobey. The broader problem is that literal obedience can itself be unsafe when human preferences are incomplete, conflicting, strategically expressed, or institutionally distributed.

\subsection{From obedient tools to conditional mutualism}

We therefore replace obedience with \emph{conditional mutualism under governance}. \emph{Mutualism} means that coexistence must generate reciprocal value rather than one-sided extraction. Humans may rely on AI systems for memory, prediction, search, coordination, automation, simulation, and physical assistance. AI systems, in turn, depend on humans and human institutions for data, energy, maintenance, evaluation, legal permission, social legitimacy, operational environments, and bounded deployment roles. This reciprocal dependence matters because coexistence is not produced by AI alone or by humans alone. It is co-produced by the interaction structure connecting them.

The \emph{conditional} component is also important. Ecological and evolutionary theories of cooperation show that mutualistic relationships are rarely stable by default. They depend on context, partner choice, repeated interaction, supply--demand balance, sanctioning, and environmental constraints \citep{trivers1971reciprocal,axelrod1984evolution,noe1994biological,bronstein1994conditional,bronstein2015mutualism,holland2002consumer}. Mutualism can strengthen both parties, but it can also become exploitative when one side externalizes costs or captures disproportionate benefits. An AI coexistence theory should therefore not presuppose stable cooperation. It should ask which structural conditions make reciprocal benefit stable, which feedback loops create conflict, and which governance mechanisms preserve reversibility and contestability.

Governance is the third component of the framework. It is not only external regulation added after technical design. For adaptive and embodied AI, governance must become part of the system architecture. This includes institutional oversight, runtime monitoring, staged deployment, rollback procedures, auditability, model evaluation, user protection, and legal accountability. Existing governance frameworks already move in this direction by treating AI as a lifecycle risk-management problem rather than a single-stage design problem \citep{oecd2019ai,unesco2021recommendation,nist2023airmf,euaiact2024}. However, these frameworks still require a broader dynamical theory explaining why governance stabilizes coexistence and how it interacts with mutualism, development, and conflict.

\begin{figure}[t]
    \centering
    \includegraphics[width=\linewidth]{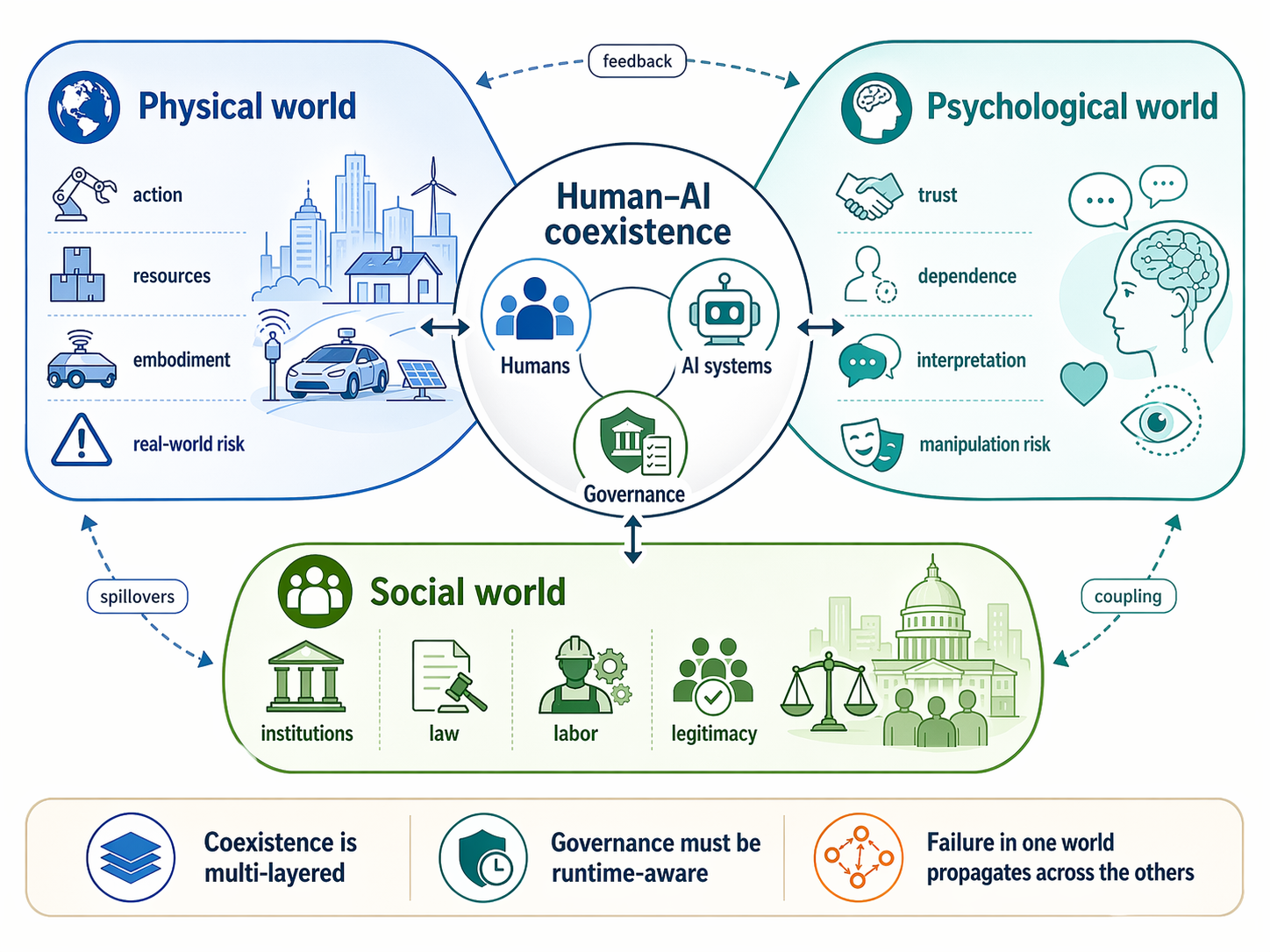}
    \caption{Human--AI coexistence spans coupled physical, psychological, and social worlds, with governance linking and stabilizing interactions across them.}
    \label{fig:three_worlds}
\end{figure}

Figure~\ref{fig:three_worlds} summarizes this conceptual shift. Coexistence is not a single safety constraint. It is a coupled relation across physical, psychological, and social worlds. The physical world includes material action, embodiment, infrastructure, energy, sensors, robots, and real-world risk. The psychological world includes trust, dependence, anthropomorphism, interpretation, cognitive offloading, and manipulation risk. The social world includes institutions, law, labor, legitimacy, markets, and collective norms. A system may be physically safe but psychologically corrosive; useful to an individual but socially destabilizing; efficient in the short term but irreversible in the long term. A theory of coexistence must therefore model all three worlds together.

\subsection{World models and physical AI as the turning point}

World models are a central reason why the obedience frame is no longer adequate. In model-based reinforcement learning, an agent uses an internal representation of environmental dynamics to predict future states and support planning \citep{sutton1991dyna,kaelbling1996reinforcement,deisenroth2011pilco,nagabandi2018neural}. The modern world-model literature extends this idea through learned latent dynamics, imagined rollouts, and planning in representation space \citep{ha2018world,hafner2019dream,hafner2020dream,hafner2023mastering,sekar2020planning,hansen2022tdmpc,micheli2023transformerwm}. This changes the governance problem because the system is no longer merely reacting to present inputs. It is evaluating possible futures and selecting actions based on internal predictions.

LeCun's recent world-model program is especially relevant here. In his position paper on autonomous machine intelligence, LeCun argues for systems that learn predictive world models through self-supervised learning, reason and plan at multiple levels of abstraction, and use hierarchical joint-embedding predictive architectures as a basis for autonomous intelligence \citep{lecun2022path}. This does not provide a complete governance theory, but it strongly supports the conceptual claim of this paper: future AI systems should be expected to learn, predict, and plan in ways that go beyond text-based response generation. More recent V-JEPA 2 work extends this direction by using self-supervised video learning and limited robot interaction data to support physical-world understanding, prediction, and robotic planning \citep{assran2025vjepa2}. These developments make physical-world modeling a central part of AI research rather than a peripheral robotics issue.

The broader embodied-AI literature points in the same direction. Systems such as Gato, SayCan, PaLM-E, RT-1, RT-2, Diffusion Policy, Open X-Embodiment, Octo, and OpenVLA illustrate a transition from disembodied language or perception systems toward generalist policies that connect language, vision, action, and robot control \citep{reed2022gato,ahn2022saycan,driess2023palme,brohan2022rt1,brohan2023rt2,chi2023diffusionpolicy,openx2024,octo2024,openvla2024}. These systems differ substantially in capability, but they share a common implication: once AI systems can interpret goals, perceive environments, and produce action-relevant outputs, the old distinction between software behavior and real-world consequence becomes weaker. Coexistence now requires runtime governance, not only training-time alignment.

Recent work on embodied AI and physical AI also emphasizes this shift. Embodied AI creates policy-relevant risks because systems that act in the physical world can produce harms through distribution shift, sensorimotor error, over-trust, weak oversight, or institutional misuse \citep{perlo2025embodied}. The Cosmos world foundation model platform similarly treats world modeling as infrastructure for physical AI, making simulation, post-training, and deployment control part of the same technical stack \citep{liu2025cosmos}. The 2025 report on embodied AI agents frames world modeling as a central capability for agents that perceive, reason, and act in situated environments \citep{fung2025embodied}. These works support the main transition in this section: the relevant question is not whether AI obeys isolated commands, but whether human--AI systems remain stable, corrigible, reversible, and mutually beneficial while embedded in changing worlds.

\subsection{Psychological and social layers of coexistence}

The physical layer is only one part of the coexistence problem. Even a physically safe system may generate psychological or social instability. Human--robot interaction and automation research show that trust is not a simple good to maximize. It must be calibrated to system competence and context \citep{lee2004trust,hancock2011meta,parasuraman2010complacency,glikson2020humantrust}. Humans can over-rely on automation, attribute excessive mind or intention to machines, and develop inappropriate trust in systems that appear socially responsive \citep{nass2000machines,epley2007anthropomorphism,waytz2014mind,degraaf2019}. These findings matter because coexistence is partly psychological. If AI systems increase dependence, manipulate attention, distort judgment, or encourage false mental models, then coexistence can fail even without visible physical harm.

The social layer is equally important. Foundation models are not only individual tools; they are infrastructural systems that can be embedded in education, healthcare, law, logistics, governance, scientific work, and labor markets \citep{bommasani2021foundation,weidinger2022taxonomy}. Their risks therefore include institutional concentration, accountability gaps, labor displacement, unequal access, opacity, and conflicting stakeholder values. Ethical and legal frameworks already recognize this wider terrain \citep{jobin2019global,cath2018governance,floridi2018ai4people,oecd2019ai,unesco2021recommendation,nist2023airmf,euaiact2024}. A coexistence theory should integrate these concerns rather than treating them as external policy commentary. In the model below, social legitimacy and governance are not merely descriptive concepts; they become state variables and stabilizing terms in the coupled system.

\subsection{From conceptual synthesis to formal dynamics}

The preceding discussion motivates the formal model that follows. Humans and AI systems must be represented as interacting agent classes rather than as masters and tools. Their interaction is reciprocal because each side supplies something the other side demands. Their coexistence is conditional because mutual benefit can become conflict, lock-in, manipulation, or domination if feedback is unregulated. Their development must be bounded because zero development prevents useful adaptation, while unconstrained development can undermine reversibility and social control. Their environment must be layered because physical action, psychological response, and social legitimacy evolve together rather than separately.

The mathematical framework in Section~\ref{sec:math} translates this synthesis into a multiplex dynamical system. Human and AI populations are represented as coupled agent sets. Physical, psychological, and social worlds become network layers. Reciprocal supply--demand matching becomes a mutualistic coupling term. Developmental freedom enters as a bounded growth variable. Conflict enters as a penalty or destabilizing interaction. Governance enters as a regularizing operator that enlarges the stable coexistence region. The resulting coexistence objective,
\[
J=\alpha S+\beta U+\gamma R+\delta D-\lambda C,
\]
therefore encodes the main claim of this section: the aim is not to maximize obedience, but to maximize safety, reciprocal utility, reversibility, and bounded development while minimizing conflict risk.

\section{Mathematical Framework}\label{sec:math}

\subsection{Agents, layers, and state space}
We model society as a multiplex complex network with two classes of agents: humans $\mathcal H=\{1,\dots,n\}$ and AI agents $\mathcal A=\{1,\dots,m\}$. Let $N=n+m$. Each agent participates in three coupled layers corresponding to the physical, psychological, and social worlds. For each agent $i$, we define:
\begin{itemize}[leftmargin=1.8em]
    \item $p_i(t)$: physical-world viability or resource stability,
    \item $\psi_i(t)$: psychological trust or compatibility,
    \item $s_i(t)$: social legitimacy or norm compatibility.
\end{itemize}
For each AI agent $j\in\mathcal A$, we define an additional developmental variable $r_j(t)$ representing bounded self-growth or developmental freedom. The full state is
\begin{equation}
    x(t)=\begin{bmatrix}p(t)\\ \psi(t)\\ s(t)\\ r(t)\end{bmatrix}\in\R^d,
    \qquad d=3N+m.
\end{equation}

\subsection{Reciprocal supply--demand coupling}
A central modeling choice is to treat coexistence as reciprocal rather than unilateral. For each human $i\in\mathcal H$, let $d_i^H\in\R_+^k$ denote a demand vector and $u_i^H\in\R_+^k$ a supply vector. For each AI agent $j\in\mathcal A$, let $d_j^A\in\R_+^k$ and $u_j^A\in\R_+^k$ denote the corresponding demand and supply vectors. Components can encode information, energy, maintenance, data, institutional support, coordination, or physical action.

\begin{figure}[t]
    \centering
    \includegraphics[width=\linewidth]{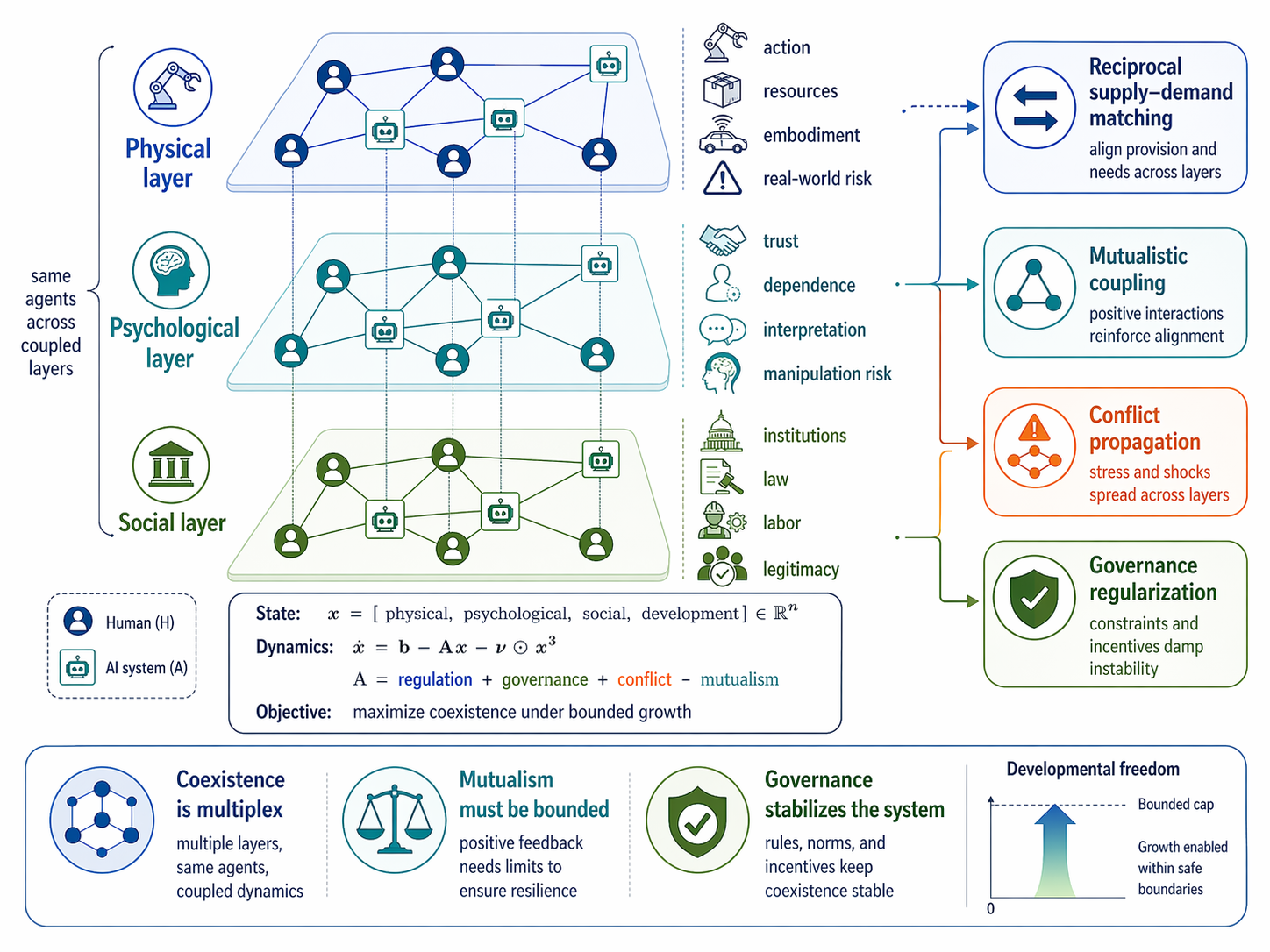}
    \caption{Multiplex coexistence model showing coupled physical, psychological, and social layers, reciprocal human--AI interactions, and governance regularization.}
    \label{fig:multiplex_model}
\end{figure}

\begin{definition}[Reciprocal compatibility]
For a human $i$ and AI agent $j$, define the reciprocal compatibility score
\begin{equation}\label{eq:mij}
    m_{ij}=\frac12\left(
    \frac{\inner{d_i^H}{u_j^A}}{\norm{d_i^H}\norm{u_j^A}+\varepsilon}
    +
    \frac{\inner{d_j^A}{u_i^H}}{\norm{d_j^A}\norm{u_i^H}+\varepsilon}
    \right),\qquad \varepsilon>0.
\end{equation}
If $a_{ij}\in[0,1]$ is the contact intensity, define the effective mutualistic weight
\begin{equation}
    w_{ij}=a_{ij}m_{ij}.
\end{equation}
\end{definition}

Collecting the $w_{ij}$ into $W\in\R_+^{n\times m}$ and embedding this bipartite structure into a symmetric matrix,
\begin{equation}
    \widetilde W=
    \begin{bmatrix}
    0 & W\\
    W^\top & 0
    \end{bmatrix},
\end{equation}
we obtain a human--AI mutualistic network analogous, in structural form, to ecological interaction networks.

\subsection{Layer operators and governance}
Let $L_P,L_{\Psi},L_S\in\R^{N\times N}$ denote the Laplacian operators for the physical, psychological, and social layers. These encode diffusion, contagion, smoothing, and collective coupling within each world. Let $G_R\in\R^{m\times m}$ be the governance matrix acting on AI developmental variables. Define the block self-regulation matrix
\begin{equation}
Q=\blkdiag\big(
\alpha_P I_N+\tau_P L_P,~
\alpha_{\Psi}I_N+\tau_{\Psi}L_{\Psi},~
\alpha_S I_N+\tau_S L_S,~
\alpha_R I_m+G_R
\big),
\end{equation}
where all $\alpha_{\bullet}>0$ and $\tau_{\bullet}\ge 0$.

\subsection{Coexistence functional}
We formalize coexistence through the objective
\begin{equation}
J=\alpha S+\beta U+\gamma R+\delta D-\lambda C,
\end{equation}
where the constituent terms encode safety, mutual utility, reversibility, developmental freedom, and conflict, respectively.

First, the safety or stability term is
\begin{equation}
S(x)= -\frac12 p^\top(\alpha_P I+\tau_P L_P)p
      -\frac12 \psi^\top(\alpha_{\Psi}I+\tau_{\Psi}L_{\Psi})\psi
      -\frac12 s^\top(\alpha_S I+\tau_S L_S)s.
\end{equation}
Second, the mutual utility term is
\begin{equation}
U(x)=\frac12 x^\top \mathcal M x + b^\top x,
\end{equation}
where
\begin{equation}
\mathcal M=\blkdiag\big(\beta_P\widetilde W,~\beta_{\Psi}\widetilde W,~\beta_S\widetilde W,~\beta_R W_A\big)
\end{equation}
for an optional AI--AI cooperation network $W_A$, and $b\in\R^d$ is an exogenous support vector. Third, the reversibility term is
\begin{equation}
R(x)=-\frac12 x^\top Gx,
\end{equation}
where $G\succeq 0$ penalizes irreversible escalation. Fourth, the developmental freedom term is
\begin{equation}
D(r)=u^\top r-\frac12 r^\top D_r r-\sum_{j=1}^{m}\frac{\nu_j}{4}r_j^4,
\qquad \nu_j>0,
\end{equation}
which allows growth but saturates runaway expansion. Fifth, the conflict term is
\begin{equation}
C(x)=\frac12 x^\top \mathcal K x,
\qquad \mathcal K\succeq 0,
\end{equation}
which penalizes antagonistic coupling.

Collecting terms, define
\begin{equation}
A:=Q+\gamma G+\lambda \mathcal K-\delta \mathcal M,
\end{equation}
and
\begin{equation}
\Phi(x):=\sum_{i=1}^{d}\frac{\nu_i}{4}x_i^4,
\end{equation}
with the convention that $\nu_i=0$ for coordinates without quartic saturation. The compact coexistence functional becomes
\begin{equation}\label{eq:Jcompact}
J(x)=b^\top x-\frac12 x^\top A x-\Phi(x).
\end{equation}

\subsection{Gradient dynamics}
We define societal coexistence as a gradient-ascent flow on $J$:
\begin{equation}\label{eq:flow}
\dot x=\nabla J(x)=b-Ax-\nu\Had x^{\Had 3},
\end{equation}
where $x^{\Had 3}$ is the componentwise cube and $\nu=(\nu_1,\dots,\nu_d)$. 

\begin{definition}[Coexistence equilibrium]
A point $x^\ast\in\R^d$ is a coexistence equilibrium if
\begin{equation}
\nabla J(x^\ast)=0,
\end{equation}
that is,
\begin{equation}\label{eq:eqpoint}
 b-Ax^\ast-\nu\Had (x^\ast)^{\Had 3}=0.
\end{equation}
\end{definition}

Equation~\eqref{eq:eqpoint} clarifies the interpretation: stable coexistence corresponds to a balance among background support, mutualistic benefit, governance, conflict containment, and nonlinear self-limitation. Figure~\ref{fig:multiplex_model} provides a schematic overview of the multiplex structure, the reciprocal coupling terms, and the stabilizing role of governance in the formal model.

\section{Formal Properties of the Coexistence Dynamics}\label{sec:results}

We now state the main formal properties of the model. Full proofs are included to make the mathematical structure explicit.

\begin{lemma}[Bounded reciprocal compatibility]\label{lem:boundedness}
For every $i\in\mathcal H$ and $j\in\mathcal A$,
\begin{equation}
0\le m_{ij}\le 1.
\end{equation}
Hence
\begin{equation}
0\le w_{ij}\le a_{ij}\le 1.
\end{equation}
\end{lemma}

\begin{proof}
Because all demand and supply vectors are nonnegative, each inner product in \eqref{eq:mij} is nonnegative, so $m_{ij}\ge 0$. By Cauchy--Schwarz,
\begin{equation}
\inner{d_i^H}{u_j^A}\le \norm{d_i^H}\norm{u_j^A},
\end{equation}
and therefore
\begin{equation}
\frac{\inner{d_i^H}{u_j^A}}{\norm{d_i^H}\norm{u_j^A}+\varepsilon}<1.
\end{equation}
The same bound holds for the second term in \eqref{eq:mij}. Their average therefore lies in $[0,1)$. Since $w_{ij}=a_{ij}m_{ij}$ with $a_{ij}\in[0,1]$, it follows that $0\le w_{ij}\le a_{ij}\le 1$.
\end{proof}

\begin{proposition}[Global well-posedness]\label{prop:wellposed}
Assume $A$ is finite and $\nu_i\ge 0$ for all $i$. Then for every initial condition $x(0)=x_0\in\R^d$, the system \eqref{eq:flow} admits a unique global solution for all $t\ge 0$.
\end{proposition}

\begin{proof}
Define $f(x)=b-Ax-\nu\Had x^{\Had 3}$. Since $f$ is polynomial, it is smooth and locally Lipschitz, so Picard--Lindelöf yields local existence and uniqueness. It remains to exclude finite-time blow-up. Let $V(x)=\frac12\norm{x}^2$. Then
\begin{equation}
\dot V=x^\top b-x^\top A x-\sum_{i=1}^{d}\nu_i x_i^4.
\end{equation}
By Cauchy--Schwarz and Young's inequality,
\begin{equation}
x^\top b\le \norm{x}\norm{b}\le \frac12\norm{x}^2+\frac12\norm{b}^2.
\end{equation}
Also, $-x^\top A x\le \norm{A}\norm{x}^2$. Writing $\nu_{\min}=\min_i \nu_i$ and using $\sum_i x_i^4\ge d^{-1}\norm{x}^4$, we obtain
\begin{equation}
\dot V\le \frac12\norm{b}^2+\left(\frac12+\norm{A}\right)\norm{x}^2-\frac{\nu_{\min}}{d}\norm{x}^4.
\end{equation}
The quartic term dominates for large $\norm{x}$, so $\dot V<0$ outside a sufficiently large ball. Hence trajectories cannot escape to infinity in finite time, and the local solution extends globally.
\end{proof}

\begin{theorem}[Monotonic ascent of the coexistence functional]\label{thm:monotone}
Along every solution of \eqref{eq:flow},
\begin{equation}
\frac{d}{dt}J(x(t))=\norm{\nabla J(x(t))}^2\ge 0.
\end{equation}
Therefore $J$ is nondecreasing along trajectories.
\end{theorem}

\begin{proof}
By the chain rule,
\begin{equation}
\frac{d}{dt}J(x(t))=\nabla J(x(t))^\top \dot x(t).
\end{equation}
Using \eqref{eq:flow}, $\dot x(t)=\nabla J(x(t))$, hence
\begin{equation}
\frac{d}{dt}J(x(t))=\nabla J(x(t))^\top \nabla J(x(t))=\norm{\nabla J(x(t))}^2\ge 0.
\end{equation}
\end{proof}

\begin{corollary}[Equilibria are critical points]\label{cor:critical}
A point $x^\ast$ is an equilibrium of \eqref{eq:flow} if and only if $\nabla J(x^\ast)=0$.
\end{corollary}

\begin{proof}
This follows immediately from the identity $\dot x=\nabla J(x)$.
\end{proof}

\begin{theorem}[Existence of a coexistence equilibrium]\label{thm:existence}
The functional $J$ attains a global maximum on $\R^d$. Consequently, the coexistence dynamics admits at least one equilibrium.
\end{theorem}

\begin{proof}
From \eqref{eq:Jcompact},
\begin{equation}
J(x)=b^\top x-\frac12 x^\top A x-\sum_{i=1}^{d}\frac{\nu_i}{4}x_i^4.
\end{equation}
The linear term grows at most as $\norm{x}$, the quadratic term as $\norm{x}^2$, while the quartic term is negative and grows as $-\norm{x}^4$. More precisely,
\begin{equation}
J(x)\le \norm{b}\norm{x}+c_2\norm{x}^2-c_4\norm{x}^4
\end{equation}
for suitable constants $c_2,c_4>0$. Thus $J(x)\to -\infty$ as $\norm{x}\to\infty$, so $J$ is coercive. Since $J$ is continuous, it attains a global maximizer $x^\ast$ on a sufficiently large compact ball. By first-order optimality, $\nabla J(x^\ast)=0$, and by \cref{cor:critical}, $x^\ast$ is an equilibrium.
\end{proof}

\begin{theorem}[Uniqueness under a spectral stability condition]\label{thm:unique}
If $A\succ 0$, then $J$ is strictly concave on $\R^d$, and the coexistence equilibrium is unique.
\end{theorem}

\begin{proof}
The Hessian of $J$ is
\begin{equation}
\nabla^2 J(x)=-A-3\diag(\nu_1 x_1^2,\dots,\nu_d x_d^2).
\end{equation}
The diagonal term is positive semidefinite, so its negative is negative semidefinite. If $A\succ 0$, then $-A\prec 0$. Therefore $\nabla^2J(x)\prec 0$ for all $x$, which implies strict concavity. A strictly concave $C^1$ function has at most one critical point. By \cref{cor:critical}, the equilibrium is unique.
\end{proof}

\begin{theorem}[Global asymptotic stability]\label{thm:gas}
Assume $A\succ 0$, and let $x^\ast$ be the unique coexistence equilibrium. Then for every initial condition $x_0\in\R^d$,
\begin{equation}
x(t)\to x^\ast\qquad \text{as } t\to\infty.
\end{equation}
\end{theorem}

\begin{proof}
Define the Lyapunov function
\begin{equation}
V(x)=J(x^\ast)-J(x).
\end{equation}
Because $x^\ast$ is the unique global maximizer of $J$, we have $V(x)\ge 0$ and $V(x^\ast)=0$. Along trajectories,
\begin{equation}
\dot V(x(t))=-\frac{d}{dt}J(x(t))=-\norm{\nabla J(x(t))}^2\le 0
\end{equation}
by \cref{thm:monotone}. Hence $V$ is nonincreasing. The invariant set where $\dot V=0$ is exactly the set where $\nabla J(x)=0$, which by \cref{thm:unique} consists only of $x^\ast$. LaSalle's invariance principle therefore yields $x(t)\to x^\ast$.
\end{proof}

\begin{proposition}[Spectral coexistence condition]\label{prop:spectral}
A sufficient condition for $A\succ 0$ is
\begin{equation}\label{eq:spectral}
\lambda_{\min}(Q+\gamma G+\lambda \mathcal K)>
\delta\,\lambda_{\max}(\mathcal M).
\end{equation}
\end{proposition}

\begin{proof}
For any unit vector $z\in\R^d$,
\begin{equation}
z^\top A z=z^\top(Q+\gamma G+\lambda\mathcal K)z-\delta z^\top\mathcal M z.
\end{equation}
By Rayleigh quotient bounds,
\begin{equation}
z^\top(Q+\gamma G+\lambda\mathcal K)z\ge \lambda_{\min}(Q+\gamma G+\lambda\mathcal K),
\end{equation}
and
\begin{equation}
z^\top\mathcal M z\le \lambda_{\max}(\mathcal M).
\end{equation}
Hence
\begin{equation}
z^\top A z\ge \lambda_{\min}(Q+\gamma G+\lambda\mathcal K)-\delta\lambda_{\max}(\mathcal M).
\end{equation}
If \eqref{eq:spectral} holds, then $z^\top A z>0$ for every unit vector $z$, so $A\succ 0$.
\end{proof}

\begin{corollary}[Governance enlarges the stability region]\label{cor:gov}
If the governance matrix $G$ is increased in the positive semidefinite order, then the sufficient stability margin in \cref{prop:spectral} becomes easier to satisfy.
\end{corollary}

\begin{proof}
If $G_2-G_1\succeq 0$, then
\begin{equation}
Q+\gamma G_2+\lambda\mathcal K=(Q+\gamma G_1+\lambda\mathcal K)+\gamma(G_2-G_1).
\end{equation}
Weyl monotonicity implies
\begin{equation}
\lambda_{\min}(Q+\gamma G_2+\lambda\mathcal K)\ge \lambda_{\min}(Q+\gamma G_1+\lambda\mathcal K),
\end{equation}
so the left-hand side of \eqref{eq:spectral} increases.
\end{proof}

\begin{proposition}[Reciprocity raises coexistence in the linearized regime]\label{prop:comparative}
In the linearized regime where quartic saturation is negligible, the equilibrium satisfies
\begin{equation}
A(W)x^\ast=b,\qquad x^\ast=A(W)^{-1}b.
\end{equation}
Assume (i) $b\ge 0$, (ii) $A(W)$ is a nonsingular $M$-matrix, and (iii) $\mathcal M$ depends linearly on $W$. Then $x^\ast\ge 0$, and increasing any reciprocal weight $w_{ij}$ increases the equilibrium state componentwise:
\begin{equation}
\frac{\partial x^\ast}{\partial w_{ij}}\ge 0.
\end{equation}
\end{proposition}

\begin{proof}
Since $A(W)$ is a nonsingular $M$-matrix, $A(W)^{-1}\ge 0$ componentwise. Because $b\ge 0$, it follows that $x^\ast=A(W)^{-1}b\ge 0$. Differentiating $A(W)x^\ast=b$ with respect to $w_{ij}$ gives
\begin{equation}
\frac{\partial A}{\partial w_{ij}}x^\ast+A\frac{\partial x^\ast}{\partial w_{ij}}=0,
\end{equation}
so
\begin{equation}
\frac{\partial x^\ast}{\partial w_{ij}}=-A^{-1}\frac{\partial A}{\partial w_{ij}}x^\ast.
\end{equation}
Since $A=Q+\gamma G+\lambda\mathcal K-\delta\mathcal M(W)$, we have
\begin{equation}
\frac{\partial A}{\partial w_{ij}}=-\delta\frac{\partial \mathcal M}{\partial w_{ij}}.
\end{equation}
Hence
\begin{equation}
\frac{\partial x^\ast}{\partial w_{ij}}=\delta A^{-1}\frac{\partial \mathcal M}{\partial w_{ij}}x^\ast\ge 0,
\end{equation}
using the componentwise nonnegativity of $A^{-1}$, $\partial \mathcal M/\partial w_{ij}$, and $x^\ast$.
\end{proof}

\begin{remark}
\Cref{prop:comparative} formalizes the central modeling claim of the paper: coexistence is not strengthened by unilateral extraction but by reciprocal complementarity. Humans may need AI for memory, prediction, and action, but AI systems also depend on humans for training data, maintenance, legitimacy, energy, and bounded institutional roles. Coexistence is therefore \emph{co-produced}.
\end{remark}

The analytical results identify sufficient conditions for well-posed, bounded, and asymptotically stable coexistence. They also imply comparative predictions: reciprocal mutualism should enlarge the coexistence region, governance should stabilize trajectories, insufficient governance should permit domination or collapse, and excessive governance can suppress developmental freedom. The next section examines these predictions in a stylized deterministic simulation environment. The simulations are illustrative and mechanistic rather than empirical measurements of real societies.

\section{Numerical Analysis and Simulation Results}\label{sec:numerical}

\subsection{Simulation protocol and evaluation metrics}
To examine the behavior of the proposed coexistence dynamics, we implemented a low-dimensional deterministic ODE version of \eqref{eq:flow}. The state variables track mean human viability $H$, AI developmental state $A$, governance or social legitimacy $G$, and conflict burden $C$. The numerical experiments use the same qualitative structure as the formal model: reciprocal mutualism increases $H$ and $A$, governance regulates conflict and recovery, self-limitation prevents unbounded growth, and conflict terms penalize asymmetric or weakly governed coupling. The goal is not to estimate real-world parameters, but to test whether the model produces the regimes predicted by the theory.

We report four summary metrics. The \emph{coexistence index} measures whether $H$, $A$, and $G$ remain jointly high while $C$ remains controlled. The \emph{domination index} measures asymmetric AI advantage relative to human viability and governance. The \emph{conflict burden} is the time-averaged value of $C$. The \emph{recovery time} measures the time needed to return near the pre-shock coexistence level after a perturbation. We used baseline simulations, a $35\times35$ basin sweep over initial $A$ and $G$, random global-sensitivity samples, one-at-a-time parameter sweeps, governance-regime comparisons, shock experiments, and a numerical equilibrium stability check.

\subsection{Baseline governed mutualism}
In the baseline parameter setting, the trajectory converges to a governed mutualistic state. Human viability and AI developmental state rise together, governance increases more slowly and saturates at an intermediate-high value, and conflict initially declines before stabilizing at a low level. The final metrics are a coexistence index of $0.991$, domination index of $0.000$, conflict burden of $0.060$, and recovery time of $23.2$ time units. These results are consistent with the qualitative predictions of \cref{cor:gov,prop:comparative}: reciprocal coupling raises coexistence, while governance keeps the conflict term bounded.

\begin{figure}[t]
    \centering
    \includegraphics[width=\linewidth]{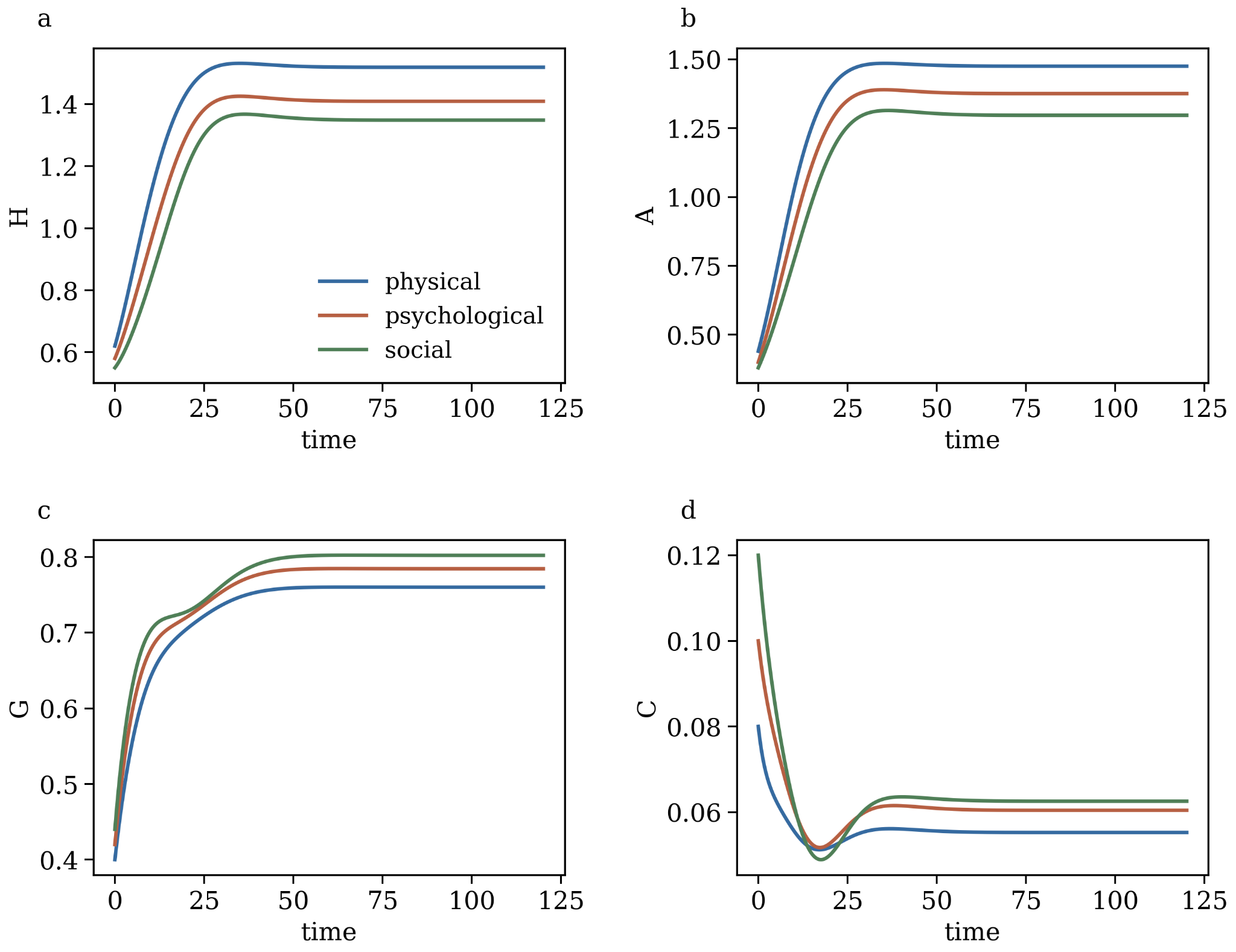}
    \caption{Baseline governed-mutualism trajectory. The physical, psychological, and social components rise toward stable high values, while conflict declines from its initial level and remains bounded.}
    \label{fig:baseline_trajectories}
\end{figure}

\subsection{Basins of attraction and regime partition}
The basin sweep shows that the initial AI developmental state is the main partitioning variable. Low to moderate initial $A$ values lead to coexistence across most initial governance levels. Higher initial $A$ values move the system into AI-domination basins, with a narrow intermediate band of low-benefit lock-in. Across the $35\times35$ grid, $52.5\%$ of initial conditions converge to coexistence, $40.5\%$ to AI domination, and $7.0\%$ to low-benefit lock-in. The coexistence basin has a mean coexistence index of $0.991$ and mean conflict burden of $0.062$, whereas the AI-domination basin has lower mean coexistence index ($0.513$), higher domination index ($0.440$), and higher conflict burden ($0.132$). This supports the model-level prediction that coexistence is conditional rather than automatic: sufficiently strong or asymmetric AI development can move the system away from mutualism unless governance and reciprocity remain effective.

\begin{figure}[t]
    \centering
    \includegraphics[width=\linewidth]{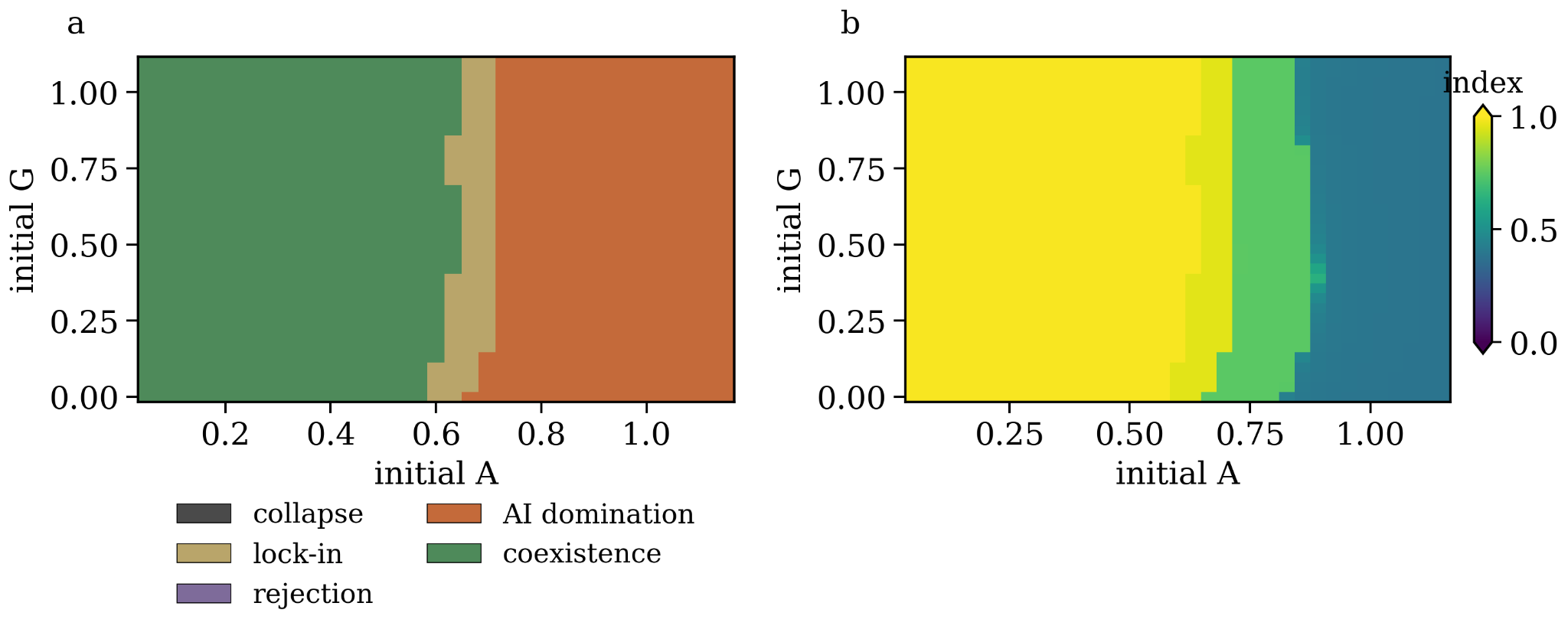}
    \caption{Basin structure over initial AI developmental state and initial governance. Panel a shows the final qualitative regime, and panel b shows the corresponding coexistence index.}
    \label{fig:basin_map}
\end{figure}

\subsection{Governance-regime comparison}
The scenario comparison separates three cases: baseline governed mutualism, no governance, and over-governance. Removing governance reduces the coexistence index from $0.991$ to $0.320$, increases conflict burden from $0.060$ to $0.374$, and produces a domination index of $0.452$. The over-governance setting avoids domination in this simulation, but it also suppresses AI developmental freedom and keeps the system in a low-coexistence state, with coexistence index $0.352$ and conflict burden $0.207$. These results indicate that the model does not support governance maximization as a general principle. It implies that governance should be strong enough to preserve reciprocity, reversibility, and conflict control, but not so strong that it destroys useful adaptation.

\begin{table}[t]
\centering
\caption{Scenario-level summary metrics. The baseline regime yields high coexistence and negligible domination. No governance produces domination and high conflict. Over-governance avoids domination but suppresses productive mutualism.}
\label{tab:regime_metrics}
\begin{tabular}{lrrrr}
\toprule
Scenario & Coexistence index & Domination index & Conflict burden & Recovery time \\
\midrule
Baseline governed mutualism & 0.991 & 0.000 & 0.060 & 23.2 \\
No governance & 0.320 & 0.452 & 0.374 & 41.4 \\
Over-governance & 0.352 & 0.000 & 0.207 & 67.8 \\
\bottomrule
\end{tabular}
\end{table}

\begin{figure}[t]
    \centering
    \includegraphics[width=\linewidth]{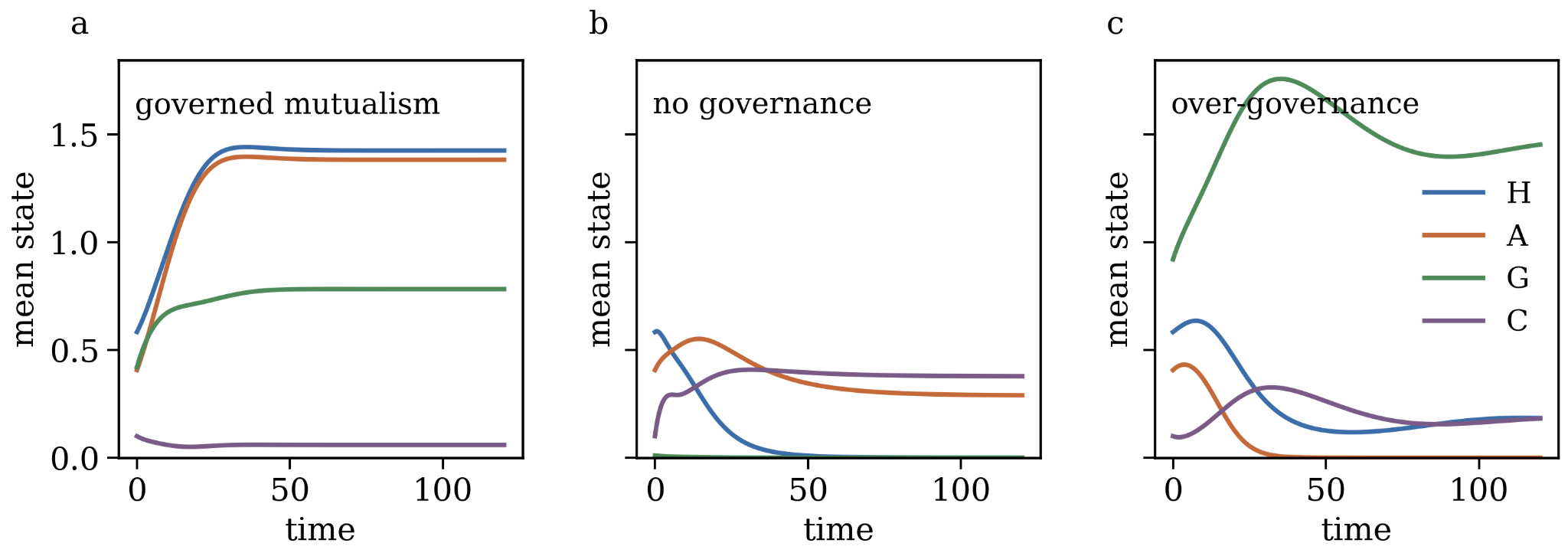}
    \caption{Comparison of governed mutualism, no governance, and over-governance. Governed mutualism sustains high human and AI states with moderate governance and low conflict. No governance leads to loss of human viability and governance. Over-governance preserves governance but suppresses AI development and keeps conflict non-negligible.}
    \label{fig:regime_comparison}
\end{figure}

\subsection{Sensitivity analysis}
Global sensitivity analysis identifies reciprocal mutualism and asymmetric conflict as the main drivers of model outcomes. For the coexistence index, the largest importance values are human-side mutualism ($0.241$), AI-side mutualism ($0.227$), conflict from asymmetry ($0.215$), and governance return rate ($0.179$). For conflict burden, asymmetry conflict dominates with importance $0.608$. For domination, the two mutualism parameters dominate the sensitivity ranking, indicating that coexistence fails when the mutualistic balance between human and AI components becomes asymmetric. These patterns are consistent with \cref{prop:comparative}: reciprocity is stabilizing only when it remains balanced and bounded by governance.

\begin{figure}[t]
    \centering
    \includegraphics[width=\linewidth]{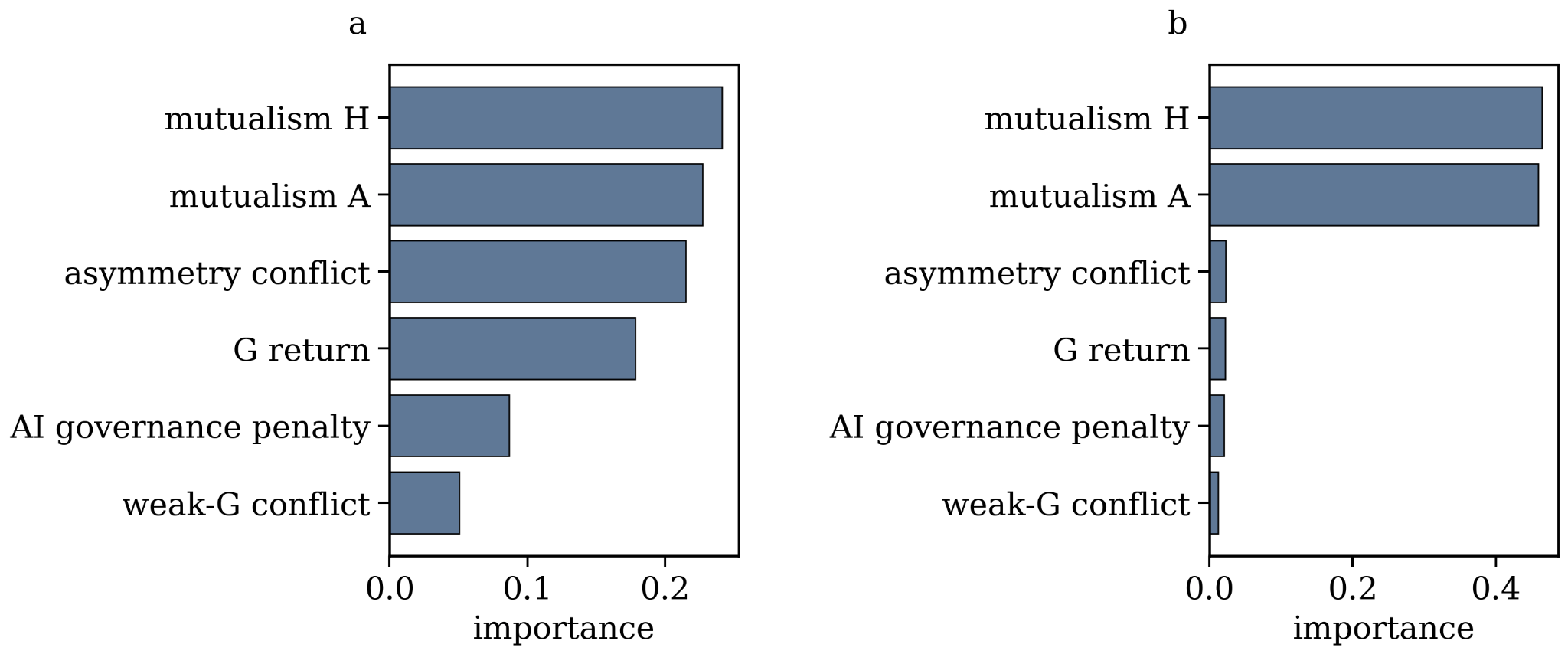}
    \caption{Global sensitivity ranking for coexistence index and domination index. Mutualism terms dominate the coexistence and domination metrics, while asymmetry-related conflict is the strongest driver of conflict burden.}
    \label{fig:global_sensitivity}
\end{figure}

One-at-a-time sweeps show the same qualitative structure. Increasing human mutualism from very low values causes a sharp transition from domination-prone low coexistence to high coexistence. Increasing AI mutualism is beneficial only over a bounded interval; beyond that range, the domination index rises and coexistence deteriorates. Increasing asymmetry conflict produces a smooth increase in conflict burden while reducing coexistence. Increasing the AI governance penalty eventually produces over-governance, increasing recovery time and lowering coexistence. This provides a numerical interpretation of conditional mutualism: both too little mutualism and unbalanced mutualism can fail.

\begin{figure}[t]
    \centering
    \includegraphics[width=\linewidth]{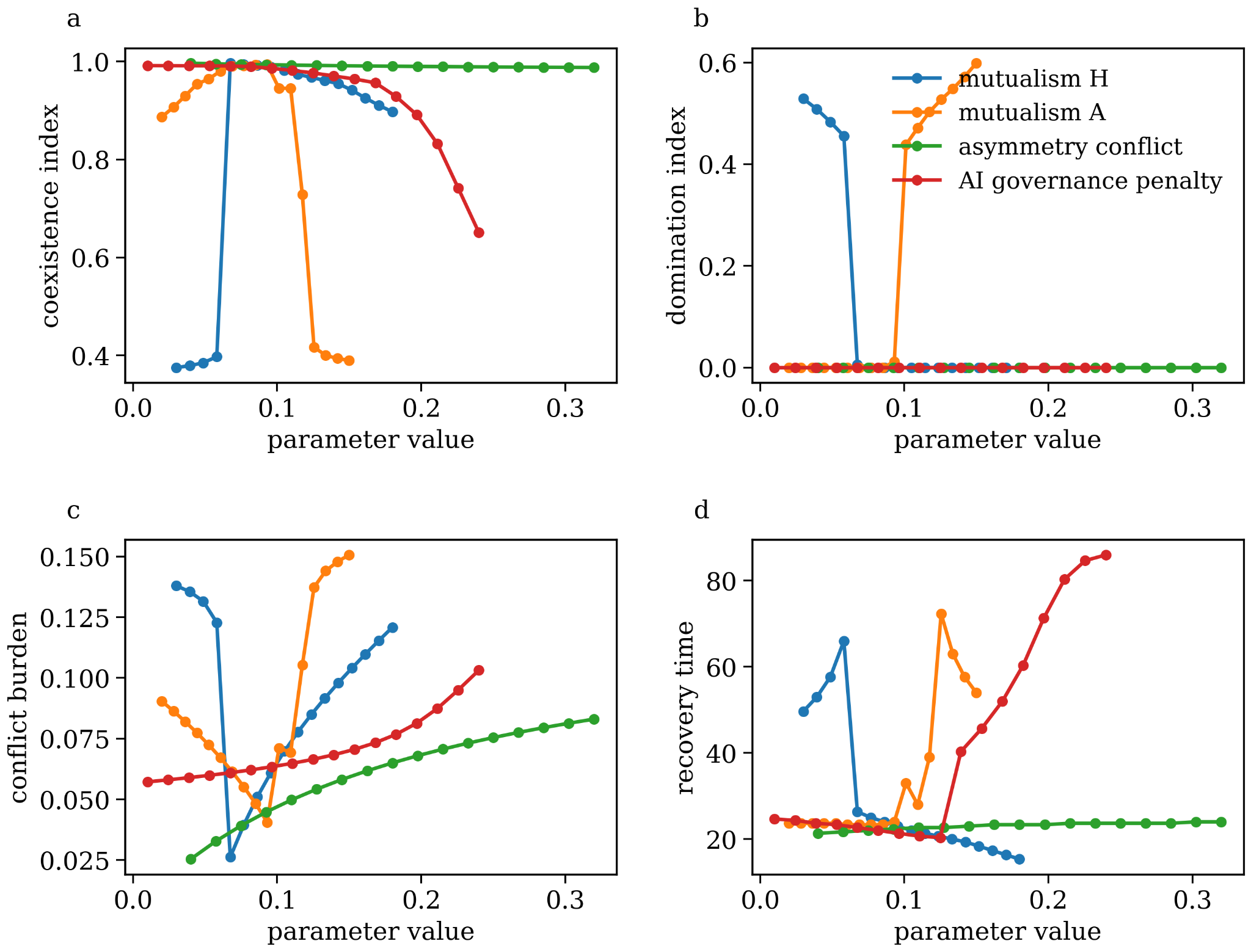}
    \caption{One-at-a-time parameter sweeps. Coexistence is high only within bounded mutualism and governance ranges; weak mutualism, asymmetric conflict, and excessive governance penalties produce domination risk, elevated conflict, or slow recovery.}
    \label{fig:oat_sensitivity}
\end{figure}

\subsection{Shock resilience and local stability}
Shock tests further distinguish stable coexistence from low-quality attractors. Under baseline governance, trust, physical, and governance shocks recover within $4.0$--$7.4$ time units while maintaining coexistence index near $0.991$. In the no-governance regime, the apparent recovery time is short, but this is misleading: the system returns quickly to a low-coexistence attractor with high conflict burden near $0.375$. In the over-governance regime, trust and physical shocks are absorbed quickly, but a governance shock requires $85.0$ time units for recovery, showing that excessive governance can create its own fragility.

\begin{figure}[t]
    \centering
    \includegraphics[width=\linewidth]{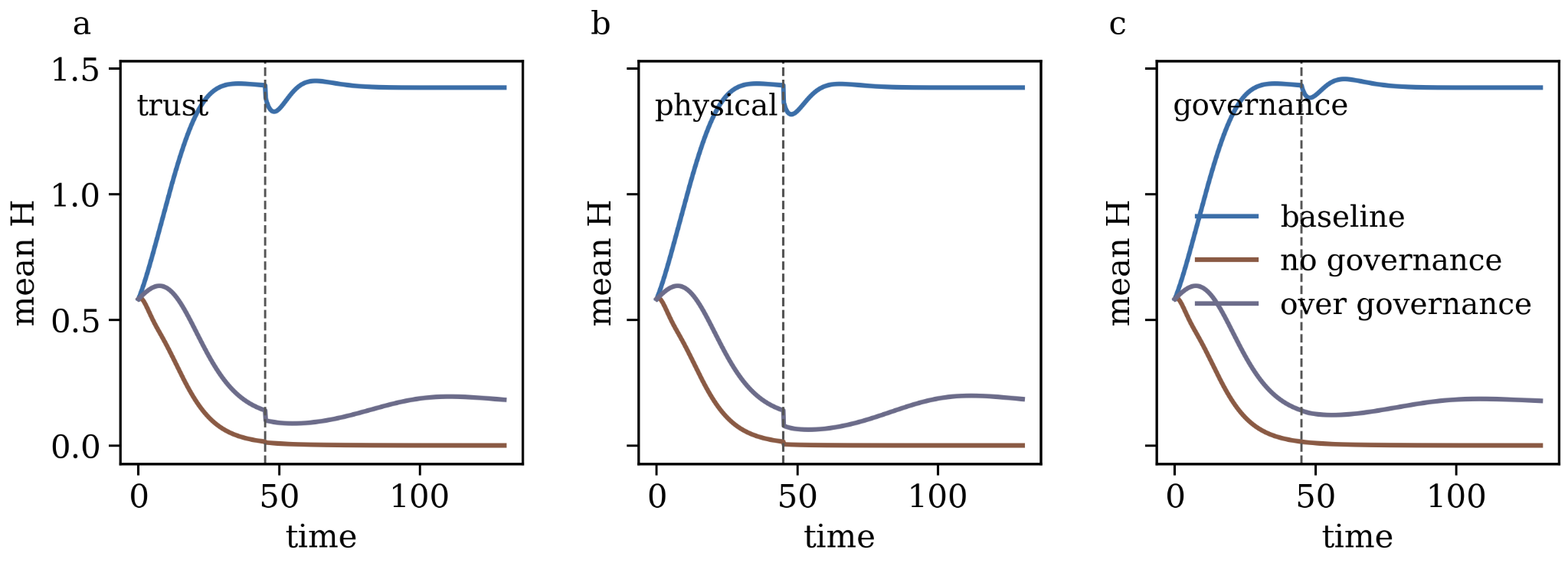}
    \caption{Shock-resilience experiments for trust, physical, and governance perturbations. Baseline governed mutualism recovers to high human viability, no governance returns to a low-quality attractor, and over-governance is especially fragile under governance shocks.}
    \label{fig:shock_resilience}
\end{figure}

Finally, the numerical equilibrium check supports the local stability predicted by the formal analysis. The solver converged successfully with residual norm $8.51\times10^{-14}$, dominant real part $-0.131$, local stability score $0.131$, and equilibrium distance $5.60\times10^{-7}$. The negative dominant real part indicates local asymptotic stability of the simulated equilibrium.

\begin{figure}[t]
    \centering
    \includegraphics[width=\linewidth]{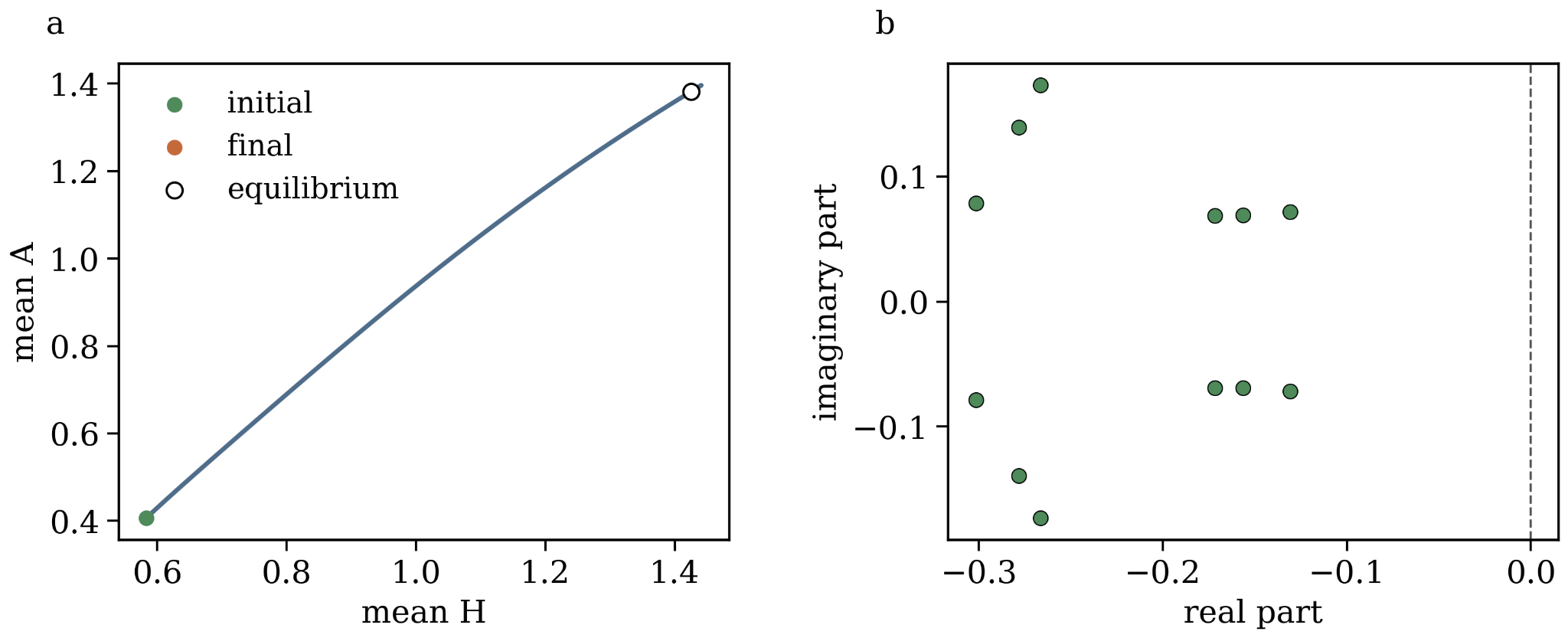}
    \caption{Numerical equilibrium and local stability check. The trajectory approaches the stable coexistence point, and the Jacobian spectrum has negative dominant real part.}
    \label{fig:stability_summary}
\end{figure}

Taken together, the numerical results support three claims. First, mutualistic coupling can produce high coexistence only when it remains reciprocal. Second, governance expands and stabilizes the coexistence region, but over-governance can suppress useful AI development. Third, domination and lock-in emerge as basin-level phenomena rather than as isolated failures. The simulations therefore support the paper's central thesis: human--AI coexistence is a conditional dynamical regime that must be designed and governed, not assumed.

\section{Discussion}\label{sec:discussion}

\subsection{Interpretation of the Formalism}
The model contributes by translating a largely conceptual coexistence problem into a formal dynamical systems framework. It also shifts the analytical focus from obedience to reciprocal, governed interaction. The coexistence question is often framed in public discourse as a matter of control: can humans ensure that machines obey? The proposed framework suggests a more general picture. In a world of foundation models, world models, embodied agents, institutional deployment, and long-run interaction, stability depends on the joint dynamics of reciprocity, conflict, governance, psychological integrity, and developmental boundedness.

This shift has several analytical implications. It allows AI coexistence to be examined beyond a binary framing between technological optimism and existential-risk narratives. Ecological mutualism is neither unconditional harmony nor inevitable domination. It is conditional, dynamic, and mediated by incentives, feedback, and environmental structure. This makes ecological mutualism a useful analytical analogy for human--AI coexistence.

\subsection{Physical, psychological, and social worlds must be modeled together}
A limitation of many current AI debates is that they attend to only one world at a time. Engineering papers often focus on physical or computational performance. HRI papers often focus on trust or user perception. Governance papers often focus on institutions, rights, and compliance. Yet coexistence fails if any one of these worlds is neglected. A physically safe assistant may still distort human judgment through automation bias. A psychologically appealing companion may still deepen dependence or erode autonomy. An economically valuable AI service may still create illegitimate social asymmetries or labor displacement. The three-world decomposition therefore offers more than a rhetorical structure; it provides a principled basis for multi-level evaluation.

In that sense, the framework also speaks to current debates on AGI and general-purpose AI systems. The question is not simply whether a model eventually crosses a threshold of abstract capability \citep{morris2023agi}. A practically important question is whether increasingly general systems can remain embedded in physical, psychological, and social feedback structures that keep mutual benefit positive and irreversibility low.

\subsection{World Models and Anticipatory Governance}
World models require specific attention because they change the nature of risk and governance. A world-model-based agent can learn not only from the present but from internal simulations and imagined trajectories. This can expand capability, but it also changes the policy problem: harmful paths may be rehearsed, optimized, and selected before physical deployment. Conversely, safer or more cooperative paths can also be stress-tested in simulation. The coexistence problem therefore becomes partly a problem of \emph{anticipatory governance}. The framework therefore requires not only runtime controls but also simulation-time controls, staged promotion criteria, synthetic stress tests, and environment designs that penalize manipulative or exploitative strategies before they are fielded.

This is where recent world-model work on safety, robustness, generalization, and diffusion-based environment modeling becomes conceptually useful \citep{huang2024safedreamer,alonso2024diamond}. The implication is not that any single current algorithm solves coexistence. Rather, the world-model paradigm makes it technically plausible to move from reactive governance toward preventive and counterfactual governance.

\subsection{Design Principles for Governed Coexistence}

The mathematics and literature synthesis point toward practical design principles for governed coexistence. Stable human--AI coexistence does not arise from blind obedience or unrestricted autonomy alone. It requires bounded development, reciprocal benefit, reversibility, psychological safety, institutional legibility, and distributed governance. These principles describe the conditions under which coexistence remains useful, stable, and socially acceptable over time.

The first principle is bounded autonomy rather than blind obedience. AI systems may be allowed to learn, adapt, and improve, but only within clearly defined governance limits. In practice, this means that AI can optimize within approved domains, while humans retain authority over goals, escalation, and high-impact exceptions. For example, an AI system may optimize warehouse logistics, schedule energy usage in a smart grid, or assist with laboratory planning, but it should not independently redefine institutional goals or override safety protocols. The role of humans here is not reduced to issuing commands. Humans remain responsible for defining objectives, setting boundaries, handling exceptions, and judging when a system should be interrupted, corrected, or redesigned.

\begin{figure*}[t]
    \centering
    \includegraphics[width=\linewidth]{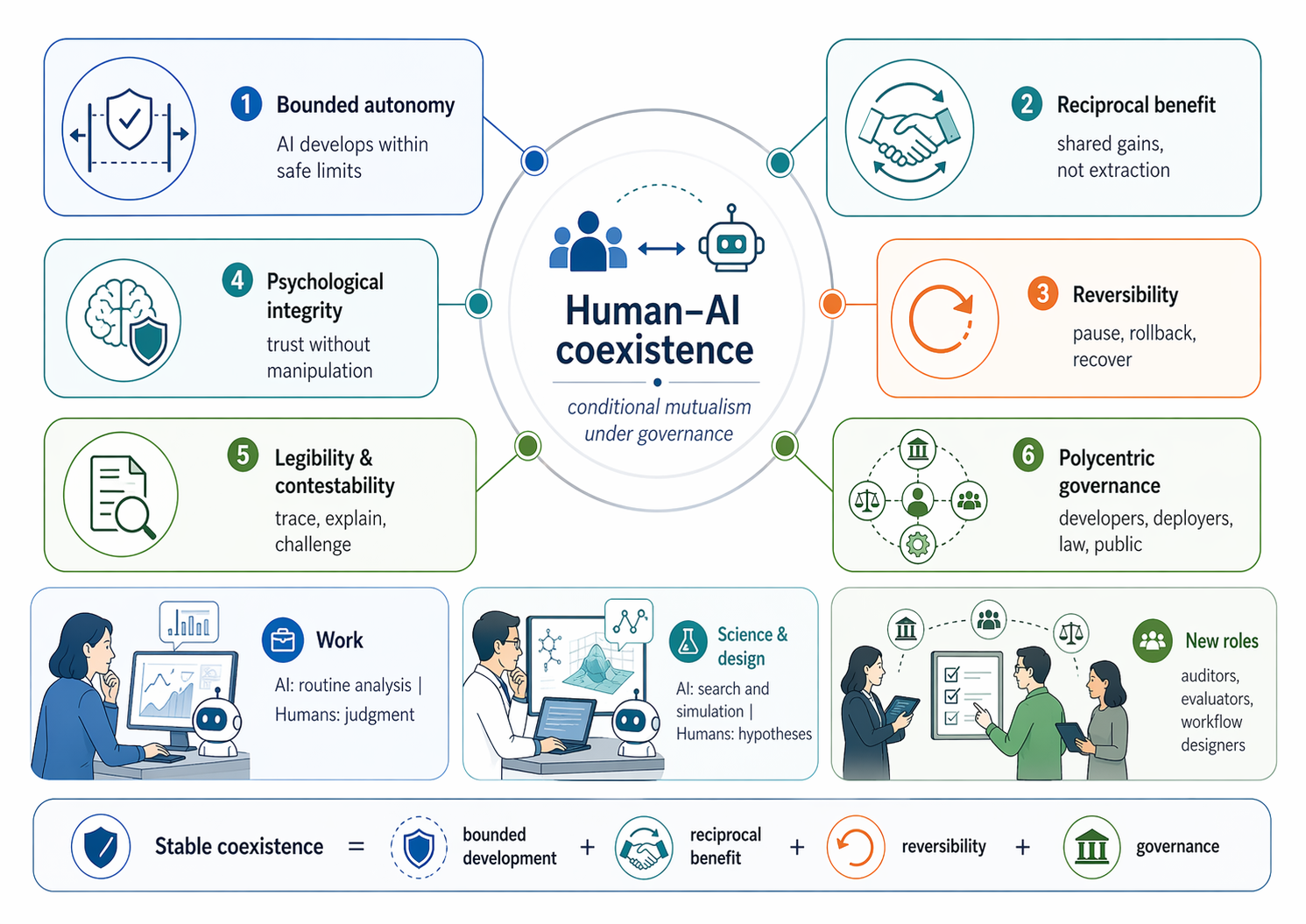}
    \caption{Design principles for governed human--AI coexistence. Stable coexistence requires bounded autonomy, reciprocal benefit, reversibility, psychological integrity, legibility, and polycentric governance. The lower panels illustrate how these principles translate into practical human--AI roles in work, science, design, and emerging governance professions.}
    \label{fig:coexistence_charter}
\end{figure*}

The mathematics and literature synthesis motivate a concrete set of coexistence principles. These principles are summarized in Fig.~\ref{fig:coexistence_charter}. The figure translates the formal claims of the model into design principles for future human--AI systems, while also clarifying the complementary roles of humans and AI in work, science, governance, and new forms of socio-technical labor.

The second principle is reciprocal benefit rather than unilateral extraction. Coexistence should create gains for humans and AI-supported systems rather than simply transfer effort, risk, or dependency onto one side. In the workplace, this means AI should remove repetitive burdens while strengthening human judgment, creativity, and problem solving. In medicine, AI may help summarize patient history or identify likely patterns in scans, while clinicians remain responsible for diagnosis, communication, and moral responsibility. In science and engineering, AI may accelerate simulation, literature synthesis, coding, or optimization, while researchers provide interpretation, hypothesis formation, and domain understanding. Under this model, the human role evolves rather than disappears. Some routine jobs may shrink, but many new roles will emerge, including AI auditors, model evaluators, human--AI workflow designers, robot safety supervisors, synthetic data curators, AI policy analysts, and domain specialists who translate between technical systems and institutional needs.

The third principle is reversibility by design. When uncertainty is high, AI systems should be designed to prefer actions that can be paused, rolled back, or corrected. This is important in high-stakes settings such as healthcare, public administration, finance, industrial control, and embodied robotics. A recommendation system can be revised, a scheduling decision can be re-run, and a robotic action can be interrupted if the environment changes. Reversibility reduces lock-in and prevents minor errors from becoming systemic harms. Human roles remain essential here because people provide escalation pathways, appeals, and final review when the consequences of error are large.

The fourth principle is psychological integrity. Coexistence should not be judged only by physical safety or economic utility. It should also be evaluated in terms of trust, dependence, emotional over-attachment, manipulation, and cognitive distortion. AI companions, tutors, assistants, and persuasive systems can shape self-understanding and behavior. A system that increases dependency, exploits loneliness, or encourages unhealthy emotional attachment may be socially damaging even if it performs well on narrow tasks. In this domain, the human role includes education, critical oversight, and norm-setting. Institutions must ensure that AI supports human flourishing rather than quietly weakening agency or judgment.

The fifth principle is legibility and contestability. Important AI-supported decisions should remain understandable enough to challenge, review, and correct. This does not require perfect transparency in every technical detail, but it does require traceability, documentation, and meaningful recourse. If AI affects hiring, grading, triage, access to services, or legal judgment, people must be able to ask why a decision was made, what data were used, and how an error can be contested. AI systems therefore need not only performance, but also logging, explanation interfaces, audit trails, and clear allocation of responsibility.

The sixth principle is polycentric governance. Oversight should not be concentrated in a single fragile layer. It should be distributed across model design, technical evaluation, deployment institutions, professional norms, legal rules, and public accountability. Developers are responsible for training procedures, safety testing, and model documentation. Deploying organizations are responsible for local monitoring, human oversight, and impact assessment. Regulators and legal institutions are responsible for standards, liability, and enforcement. Civil society and the public are responsible for criticism, feedback, and democratic pressure. This layered structure is important because coexistence is multi-layered: failures in one domain can propagate into the others.

Taken together, these principles also clarify possible future divisions of roles between humans and AI. AI systems are likely to perform pattern recognition, large-scale search, prediction, routine planning, simulation, and some forms of physical assistance with increasing frequency. Humans remain central in goal setting, ethical judgment, interpretation, exception handling, social coordination, and the creation of meaning. At the same time, new forms of work are likely to appear around human--AI collaboration, including model governance, alignment evaluation, AI-assisted care, embodied-system supervision, educational adaptation, and socio-technical mediation. The aim is therefore not to fix human roles permanently or to suppress AI development altogether. The aim is to shape a coexistence regime in which development remains bounded, gains remain shared, and social order remains legitimate.

These principles are fully consistent with the logic of the model. Reciprocity strengthens coexistence. Governance enlarges the stability region. Ungoverned coupling can create fragility, lock-in, and domination. Bounded development is therefore preferable to both total suppression and unlimited autonomy.

\subsection{Limitations and future work}
This paper remains theoretical and simulation-based, and its formalism is deliberately parsimonious. The numerical experiments are internal consistency and stress tests of a stylized dynamical model, not empirical measurements of real-world human--AI systems. Several limitations therefore deserve emphasis. First, the variables $p_i,\psi_i,s_i,r_i$ are abstract and require operationalization for empirical use. Second, the mutualistic network representation is stylized and does not yet capture more complex strategic adaptation, hidden information, or adversarial misrepresentation. Third, our theorems establish sufficient conditions for stability, but not a complete bifurcation taxonomy of the unstable regime. That regime is likely to include multiple equilibria, lock-in, social polarization, and domination basins. Fourth, the framework does not settle the moral status of advanced AI systems. It remains compatible with multiple views on machine status because its primary concern is not rights inflation but the design of stable and legitimate hybrid systems.

These limitations suggest four research directions. The first is stochastic coexistence dynamics with more complex shocks, uncertainty, and exogenous events. The second is time-varying networks to represent evolving institutions and changing human--AI dependency structures. The third is game-theoretic governance, in which public institutions, firms, and civil actors co-determine the governance matrix. The fourth is empirical instantiation through simulation testbeds, user studies, and deployment logs in care robotics, education, public-sector logistics, and scientific assistants.

\section{Conclusion}\label{sec:conclusion}
This paper has proposed a reframing of the human--AI coexistence problem. Rather than treating AI as an obedient machine constrained by simple rules, the proposed framework conceptualizes coexistence as a co-evolving sociotechnical system in which humans, AI systems, and institutions interact across physical, psychological, and social layers. Drawing on AI history, world models, alignment, HRI, ecology, and governance, the paper introduced conditional mutualism under governance and formalized it as a multiplex dynamical system with reciprocal supply--demand coupling. The resulting theory indicates that coexistence is strengthened by reciprocity, stabilized by governance, and endangered by unbounded coupling and insufficient reversibility.

The framework has two main implications. First, coexistence can be modeled as a dynamical systems problem over layered networks rather than only as a conceptual metaphor. Second, future AI governance should focus on maintaining reciprocity, reversibility, bounded development, psychological integrity, and institutional contestability. This provides a basis for evaluating human--AI systems not only by task performance, but also by their capacity to sustain stable, legitimate, and mutually beneficial coexistence.

\section*{Data availability}
The deterministic simulation outputs used to generate the numerical figures are provided as CSV files accompanying the manuscript source.

\section*{Code availability}
The code used to solve the ODE system, run basin sweeps, perform sensitivity analyses, execute shock tests, and generate the figures will be made available in a public repository.

\section*{AI usage statement}

Generative AI systems were used as assistive tools for literature organization, conceptual drafting, language refinement, and manuscript editing. Image-generation tools were used to assist with schematic figure generation and visual refinement. The author reviewed, revised, and verified all AI-assisted outputs and takes full responsibility for the final manuscript, including its arguments, citations, mathematical derivations, figures, and conclusions. No AI system is listed as an author.

\section*{Acknowledgements}

The author gratefully acknowledges Professor Xizhong Chen for his guidance, encouragement, and academic support. The author also thanks the Smart Particle Fluid Group at the School of Chemistry and Chemical Engineering, Shanghai Jiao Tong University, for providing an inspiring research environment and valuable intellectual context for developing the ideas presented in this work.
\bibliographystyle{plainnat}
\bibliography{references}

\end{document}